\documentclass[12pt]{article}

\usepackage{axodraw}

\makeatletter

\def\paragraph{\@startsection{paragraph}{4}{\z@}{+2.00ex plus
 +1ex minus +.2ex}{1.5ex plus .2ex}{\it\normalsize}}

\def\section{\@startsection {section}{1}{\z@}{+3.0ex plus +1ex minus
  +.2ex}{2.3ex plus .2ex}{\normalsize\bf\boldmath}}
\def\subsection{\@startsection{subsection}{2}{\z@}{+2.5ex plus +1ex
minus +.2ex}{1.5ex plus .2ex}{\normalsize\bf\boldmath}}
\def\subsubsection{\@startsection{subsubsection}{3}{\z@}{+3.25ex plus
 +1ex minus +.2ex}{1.5ex plus .2ex}{\normalsize\it}}

\expandafter\ifx\csname mathrm\endcsname\relax\def\mathrm#1{{\rm #1}}\fi

\renewcommand{\theequation}{\thesection.\arabic{equation}}
\newcounter{saveeqn}

\@addtoreset{equation}{section}

\newcount\@tempcntc
\def\@citex[#1]#2{\if@filesw\immediate\write\@auxout{\string\citation{#2}}\fi
  \@tempcnta\z@\@tempcntb\m@ne\def\@citea{}\@cite{\@for\@citeb:=#2\do
    {\@ifundefined
       {b@\@citeb}{\@citeo\@tempcntb\m@ne\@citea
        \def\@citea{,\penalty\@m\ }{\bf ?}\@warning
       {Citation `\@citeb' on page \thepage \space undefined}}%
    {\setbox\z@\hbox{\global\@tempcntc0\csname
b@\@citeb\endcsname\relax}%
     \ifnum\@tempcntc=\z@ \@citeo\@tempcntb\m@ne
       \@citea\def\@citea{,\penalty\@m}
       \hbox{\csname b@\@citeb\endcsname}%
     \else
      \advance\@tempcntb\@ne
      \ifnum\@tempcntb=\@tempcntc
      \else\advance\@tempcntb\m@ne\@citeo
      \@tempcnta\@tempcntc\@tempcntb\@tempcntc\fi\fi}}\@citeo}{#1}}

\def\@citeo{\ifnum\@tempcnta>\@tempcntb\else\@citea
  \def\@citea{,\penalty\@m}%
  \ifnum\@tempcnta=\@tempcntb\the\@tempcnta\else
   {\advance\@tempcnta\@ne\ifnum\@tempcnta=\@tempcntb \else
\def\@citea{--}\fi
    \advance\@tempcnta\m@ne\the\@tempcnta\@citea\the\@tempcntb}\fi\fi}

\def\nl{\nonumber\\}

\def\nln{\nonumber\\*[-1ex]\phantom{\fbox{\rule{0em}{2ex}}}}

\newcommand{\lsim}
{\mathrel{\raisebox{-.3em}{$\stackrel{\displaystyle <}{\sim}$}}}
\newcommand{\gsim}
{\mathrel{\raisebox{-.3em}{$\stackrel{\displaystyle >}{\sim}$}}}
\def\asymp#1%
{\mathrel{\raisebox{-.4em}{$\widetilde{\scriptstyle #1}$}}}

\def\Nequal#1%
{\mathrel{\raisebox{-.5em}{$\stackrel{=}{\scriptstyle\rm#1}$}}}
\newcommand{\dsl}[1]{\not \hspace{-0.7mm}#1}
\def\dsl{\mathpalette\make@slash}
\def\make@slash#1#2{\setbox\z@\hbox{$#1#2$}%
  \hbox to 0pt{\hss$#1/$\hss\kern-\wd0}\box0}

\def\beq{\begin{equation}}
\def\eeq{\end{equation}}
\def\beqar{\begin{eqnarray}}
\def\eeqar{\end{eqnarray}}
\def\barr#1{\begin{array}{#1}}
\def\earr{\end{array}}
\def\bfi{\begin{figure}}
\def\efi{\end{figure}}
\def\btab{\begin{table}}
\def\etab{\end{table}}
\def\bce{\begin{center}}
\def\ece{\end{center}}
\def\nn{\nonumber}

\def\text{\textstyle}

\def\ga{\gamma}
\def\de{\delta}
\def\veps{\varepsilon}
\def\la{\lambda}

\def\refeq#1{\mbox{(\ref{#1})}}

\def\refse#1{\mbox{Sect.~\ref{#1}}}

\def\refapp#1{\mbox{App.~\ref{#1}}}

\def\citere#1{\mbox{Ref.~\cite{#1}}}
\def\citeres#1{\mbox{Refs.~\cite{#1}}}

\def\solid{\raise.9mm\hbox{\protect\rule{1.1cm}{.2mm}}}
\def\dash{\raise.9mm\hbox{\protect\rule{2mm}{.2mm}}\hspace*{1mm}}

\newcommand{\TeV}{\unskip\,\mathrm{TeV}}

\def\mathswitchr#1{\relax\ifmmode{\mathrm{#1}}\else$\mathrm{#1}$\fi}

\newcommand{\PW}{\mathswitchr W}
\newcommand{\PZ}{\mathswitchr Z}

\newcommand{\PH}{\mathswitchr H}

\newcommand{\Pf}{\mathswitchr f}

\newcommand{\Pt}{\mathswitchr t}

\newcommand{\Pep}{\mathswitchr {e^+}}
\newcommand{\Pem}{\mathswitchr {e^-}}

\newcommand{\PWpm}{\mathswitchr {W^\pm}}

\def\mathswitch#1{\relax\ifmmode#1\else$#1$\fi}

\newcommand{\MW}{\mathswitch {M_\PW}}

\newcommand{\MZ}{\mathswitch {M_\PZ}}
\newcommand{\MH}{\mathswitch {M_\PH}}

\newcommand{\Mt}{\mathswitch {m_\Pt}}

\newcommand{\Mfl}{\mathswitch {m_{f\ne\Pt}}}

\newcommand{\thw}{\mathswitch{\theta_{\mathrm{w}}}}

\newcommand{\scrs}{\scriptscriptstyle}
\newcommand{\sw}{\mathswitch {s_{\scrs\PW}}}
\newcommand{\cw}{\mathswitch {c_{\scrs\PW}}}

\newcommand{\ew}{\mathrm{ew}}

\newcommand{\cew}{C^{\ew}}

\def\ie{i.e.\ }
\def\eg{e.g.\ }

\newcommand{\etal}{{\it et al.}}

\newcommand{\ord}{{\cal O}}

\newcommand{\LA}{\stackrel{\NLLad}{=}}

\newcommand{\SUD}{\stackrel{\mathrm{Sud}}{=}}
\newcommand{\EIK}{\stackrel{\mathrm{eik}}{=}}
\newcommand{\NLLad}{\mathswitchr{NLL_{a}}}

\newcommand{\SU}{\mathrm{SU}}
\newcommand{\U}{\mathrm{U}}
\newcommand{\SUtwo}{\mathrm{SU}(2)}
\newcommand{\Uone}{\mathrm{U}(1)}

\newcommand{\rL}{\mathrm{L}}
\newcommand{\rC}{\mathrm{C}}
\newcommand{\rY}{\mathrm{Y}}

\newcommand{\NLL}{\mathrm{NLL}}

\newcommand{\ri}{\mathrm{i}}

\newcommand{\rd}{{\mathrm{d}}}

\newcommand{\elm}{\mathrm{em}}
\newcommand{\EW}{\mathrm{EW}}
\newcommand{\sew}{\mathrm{sew}}
\newcommand{\sem}{\mathrm{sem}}

\newcommand{\M}{{\cal {M}}}

\def\sgn{\mathop{\mathrm{sgn}}\nolimits}

\def\I{S}
\def\Ione{S}
\def\Eone{E}
\def\Rone{R}
 
\newcommand{\symbdiag}{D}
 \newcommand{\diagtwoL}[2]{\symbdiag_{2{\mathrm L},#1}^{#2}}
 \newcommand{\diagtwoC}[2]{\symbdiag_{2{\mathrm C},#1}^{#2}}
 \newcommand{\diagtwoY}[2]{\symbdiag_{2{\mathrm Y},#1}^{#2}}
 \newcommand{\diagthreeL}[2]{\symbdiag_{3{\mathrm L},#1}^{#2}}
 \newcommand{\diagthreeY}[2]{\symbdiag_{3{\mathrm Y},#1}^{#2}}
 \newcommand{\diagfourL}[2]{\symbdiag_{4{\mathrm L},#1}^{#2}}

\newcommand{\leg}[1]{{#1}}

\def\draftdate{\relax}
\def\mda{\relax}
\def\mua{\relax}
\def\mla{\relax}
\def\draft{
\def\thtystars{******************************}
\def\sixtystars{\thtystars\thtystars}
\typeout{}
\typeout{\sixtystars**}
\typeout{* Draft mode!
         For final version remove \protect\draft\space in source file *}
\typeout{\sixtystars**}
\typeout{}
\def\draftdate{\today}
\def\mua{\marginpar[\boldmath\hfil$\uparrow$]%
                   {\boldmath$\uparrow$\hfil}%
                    \typeout{marginpar: $\uparrow$}\ignorespaces}
\def\mda{\marginpar[\boldmath\hfil$\downarrow$]%
                   {\boldmath$\downarrow$\hfil}%
                    \typeout{marginpar: $\downarrow$}\ignorespaces}
\def\mla{\marginpar[\boldmath\hfil$\rightarrow$]%
                   {\boldmath$\leftarrow $\hfil}%
                    \typeout{marginpar: $\leftrightarrow$}\ignorespaces}
\def\Mua{\marginpar[\boldmath\hfil$\Uparrow$]%
                   {\boldmath$\Uparrow$\hfil}%
                    \typeout{marginpar: $\Uparrow$}\ignorespaces}
\def\Mda{\marginpar[\boldmath\hfil$\Downarrow$]%
                   {\boldmath$\Downarrow$\hfil}%
                    \typeout{marginpar: $\Downarrow$}\ignorespaces}
\def\Mla{\marginpar[\boldmath\hfil$\Rightarrow$]%
                   {\boldmath$\Leftarrow $\hfil}%
                    \typeout{marginpar: $\Leftrightarrow$}\ignorespaces}

\overfullrule 5pt
\oddsidemargin -15mm
\marginparwidth 29mm
}

\makeatletter

\def\eqnarray{\stepcounter{equation}\let\@currentlabel=\theequation
\global\@eqnswtrue
\global\@eqcnt\z@\tabskip\@centering\let\\=\@eqncr
$$\halign to \displaywidth\bgroup\hskip\@centering
  $\displaystyle\tabskip\z@{##}$\@eqnsel&\global\@eqcnt\@ne
  \hskip 2\arraycolsep \hfil${##}$\hfil
  &\global\@eqcnt\tw@ \hskip 2\arraycolsep $\displaystyle\tabskip\z@{##}$\hfil
   \tabskip\@centering&\llap{##}\tabskip\z@\cr}

\makeatother

\begin{document}
\thispagestyle{empty} 

\newcommand{\smalldiagramonetwoL}{
\begin{picture}(69.75,60.45)(-16.275,-30.225)
\Line(0.,0.)(46.5,23.25)
\Line(0.,0.)(46.5,-23.25)
\Photon(27.9,13.95)(27.9,-13.95){1.395}{3.5}
\Vertex(27.9,13.95){2}
\Vertex(27.9,-13.95){2}
\Text(48.825,24.4125)[l]{$\leg{j}$}
\Text(48.825,-24.4125)[l]{$\leg{k}$}
\Text(31.1931,0.)[l]{$a$}
\Text(18.3794,9.72865)[br]{}
\Text(18.6,-9.3)[tr]{}
\blob
\end{picture}
}

\newcommand{\smalldiagramtwoL}{
\begin{picture}(69.75,60.45)(-16.275,-30.225)
\Line(0.,0.)(46.5,23.25)
\Line(0.,0.)(46.5,-23.25)
\Photon(37.2,18.6)(37.2,-18.6){1.395}{3.5}
\Photon(18.6,9.3)(18.6,-9.3){1.395}{2}
\Vertex(37.2,18.6){2}
\Vertex(37.2,-18.6){2}
\Vertex(18.6,9.3){2}
\Vertex(18.6,-9.3){2}
\Text(48.825,24.4125)[l]{$\leg{j}$}
\Text(48.825,-24.4125)[l]{$\leg{k}$}
\Text(41.5909,0.)[l]{$a$}
\Text(20.7954,0.)[l]{$b$}
\Text(27.5691,14.593)[br]{}
\Text(27.9,-13.95)[tr]{}
\Text(13.6117,7.61406)[br]{}
\Text(13.95,-6.975)[tr]{}
\blob
\end{picture}
}

\newcommand{\smalldiagramtwoC}{
\begin{picture}(69.75,60.45)(-16.275,-30.225)
\Line(0.,0.)(46.5,23.25)
\Line(0.,0.)(46.5,-23.25)
\Photon(37.2,18.6)(18.6,-9.3){-1.395}{3}
\Photon(37.2,-18.6)(18.6,9.3){-1.395}{3}
\Vertex(37.2,18.6){2}
\Vertex(37.2,-18.6){2}
\Vertex(18.6,9.3){2}
\Vertex(18.6,-9.3){2}
\Text(48.825,24.4125)[l]{$\leg{j}$}
\Text(48.825,-24.4125)[l]{$\leg{k}$}
\Text(33.2131,6.23127)[l]{$a$}
\Text(33.2131,-6.23127)[l]{$b$}
\Text(27.5691,14.593)[br]{}
\Text(27.9,-13.95)[tr]{}
\Text(13.6117,7.61406)[br]{}
\Text(13.95,-6.975)[tr]{}
\blob
\end{picture}
}

\newcommand{\smalldiagramtwoY}{
\begin{picture}(69.75,60.45)(-16.275,-30.225)
\Line(0.,0.)(46.5,23.25)
\Line(0.,0.)(46.5,-23.25)
\Photon(37.2,18.6)(31.1931,0.){1.395}{2}
\Photon(18.6,9.3)(31.1931,0.){1.395}{2}
\Photon(31.1931,0.)(31.1931,-15.5966){1.395}{2}
\Vertex(37.2,18.6){2}
\Vertex(18.6,9.3){2}
\Vertex(31.1931,0.){2}
\Vertex(31.1931,-15.5966){2}
\Text(48.825,24.4125)[l]{$\leg{j}$}
\Text(48.825,-24.4125)[l]{$\leg{k}$}
\Text(37.9484,8.9584)[l]{$a$}
\Text(35.4185,-8.36117)[l]{$b$}
\Text(23.0875,3.7798)[tr]{$c$}
\Text(27.5691,14.593)[br]{}
\Text(20.925,-10.4625)[tr]{}
\Text(13.6117,7.61406)[br]{}
\blob
\end{picture}
}

\newcommand{\smalldiagramtwoYinv}{
\begin{picture}(69.75,60.45)(-16.275,-30.225)
\Line(0.,0.)(46.5,23.25)
\Line(0.,0.)(46.5,-23.25)
\Photon(37.2,-18.6)(31.1931,0.){1.395}{2}
\Photon(18.6,-9.3)(31.1931,0.){1.395}{2}
\Photon(31.1931,0.)(31.1931,15.5966){1.395}{2}
\Vertex(37.2,-18.6){2}
\Vertex(18.6,-9.3){2}
\Vertex(31.1931,0.){2}
\Vertex(31.1931,15.5966){2}
\Text(48.825,24.4125)[l]{$\leg{j}$}
\Text(48.825,-24.4125)[l]{$\leg{k}$}
\Text(39.4663,-9.31673)[l]{$a$}
\Text(36.9364,8.71951)[l]{$b$}
\Text(23.6354,-3.64479)[br]{$c$}
\Text(27.5691,-14.593)[br]{}
\Text(20.925,10.4625)[tr]{}
\Text(13.6117,-7.61406)[br]{}
\blob
\end{picture}
}

\newcommand{\smalldiagramthreeL}{
\begin{picture}(69.75,60.45)(-16.275,-30.225)
\Line(0.,0.)(46.5,23.25)
\Line(0.,0.)(46.5,-23.25)
\Line(0.,0.)(51.9886,0.)
\Photon(39.525,19.7625)(39.525,0.){1.395}{2.5}
\Photon(26.505,-13.2525)(26.505,0.){-1.395}{2}
\Vertex(39.525,19.7625){2}
\Vertex(26.505,-13.2525){2}
\Vertex(39.525,0.){2}
\Vertex(26.505,0.){2}
\Text(48.825,24.4125)[l]{$\leg{k}$}
\Text(54.588,0.)[l]{$\leg{j}$}
\Text(48.825,-24.4125)[l]{$\leg{l}$}
\Text(43.2319,9.15318)[l]{$a$}
\Text(29.9939,-8.56608)[l]{$b$}
\Text(33.2727,0.)[b]{}
\Text(22.9743,12.1608)[br]{}
\Text(16.275,-8.1375)[tr]{}
\Text(18.196,0.)[b]{}
\blob
\end{picture}
}

\newcommand{\smalldiagramthreeLinv}{
\begin{picture}(69.75,60.45)(-16.275,-30.225)
\Line(0.,0.)(46.5,23.25)
\Line(0.,0.)(46.5,-23.25)
\Line(0.,0.)(51.9886,0.)
\Photon(26.505,13.2525)(26.505,0.){1.395}{2}
\Photon(39.525,-19.7625)(39.525,0.){-1.395}{2.5}
\Vertex(26.505,13.2525){2}
\Vertex(39.525,-19.7625){2}
\Vertex(26.505,0.){2}
\Vertex(39.525,0.){2}
\Text(48.825,24.4125)[l]{$\leg{k}$}
\Text(54.588,0.)[l]{$\leg{j}$}
\Text(48.825,-24.4125)[l]{$\leg{l}$}
\Text(43.0082,-10.1528)[l]{$b$}
\Text(30.3587,7.16672)[l]{$a$}
\Text(33.2727,0.)[t]{}
\Text(22.9743,12.1608)[br]{}
\Text(16.275,-8.1375)[tr]{}
\Text(18.196,0.)[t]{}
\blob
\end{picture}
}

\newcommand{\smalldiagramthreeY}{
\begin{picture}(69.75,60.45)(-16.275,-30.225)
\Line(0.,0.)(46.5,23.25)
\Line(0.,0.)(46.5,-23.25)
\Line(0.,0.)(51.7652,-4.81397)
\Photon(32.55,16.275)(36.0405,5.04561){1.395}{2}
\Photon(27.9,-13.95)(36.0405,5.04561){-1.395}{3}
\Photon(45.0357,-4.18815)(36.0405,5.04561){1.395}{2}
\Vertex(32.55,16.275){2}
\Vertex(45.0357,-4.18815){2}
\Vertex(27.9,-13.95){2}
\Vertex(36.0405,5.04561){2}
\Text(48.825,24.4125)[l]{$\leg{j}$}
\Text(54.3535,-5.05467)[l]{$\leg{k}$}
\Text(48.825,-24.4125)[l]{$\leg{l}$}
\Text(37.2341,11.5738)[l]{$a$}
\Text(44.1219,2.45738)[l]{$b$}
\Text(32.4934,-9.27993)[l]{$c$}
\Text(18.6,-9.3)[tr]{}
\Text(20.6769,10.9447)[br]{}
\Text(20.7954,0.)[b]{}
\blob
\end{picture}
}

\newcommand{\smalldiagramfourL}{
\begin{picture}(69.75,60.45)(-16.275,-30.225)
\Line(0.,0.)(46.5,23.25)
\Line(0.,0.)(46.5,-23.25)
\Line(0.,0.)(51.7652,4.81397)
\Line(0.,0.)(51.7652,-4.81397)
\Photon(37.2,18.6)(41.4122,3.85118){1.395}{2}
\Photon(37.2,-18.6)(41.4122,-3.85118){-1.395}{2}
\Vertex(37.2,18.6){2}
\Vertex(37.2,-18.6){2}
\Vertex(41.4122,3.85118){2}
\Vertex(41.4122,-3.85118){2}
\Text(48.825,24.4125)[l]{$\leg{j}$}
\Text(54.3535,5.05467)[l]{$\leg{k}$}
\Text(54.3535,-5.05467)[l]{$\leg{l}$}
\Text(48.825,-24.4125)[l]{$\leg{m}$}
\Text(42.4914,12.1353)[l]{$a$}
\Text(42.4914,-12.1353)[l]{$b$}
\blob
\end{picture}
}

\newcommand{\blob}{
\Vertex(-15.9138,3.75675){0.8}
\Vertex(-16.3512,-0.00758122){0.8}
\Vertex(-15.9138,-3.75675){0.8}
\Line(0.,0.)(-19.5,-9.75)
\Line(0.,0.)(-19.5,9.75)
\GCirc(0.,0.){10.9008}{0.5}
}

\newcommand{\eikdiagramonetwoL}{
\begin{picture}(97.5,84.5)(-22.75,-42.25)
\Line(0.,0.)(65.,32.5)
\Line(0.,0.)(65.,-32.5)
\Photon(39.,19.5)(39.,-19.5){1.95}{3.5}
\Vertex(39.,19.5){2}
\Vertex(39.,-19.5){2}
\Text(68.25,34.125)[l]{$\leg{j}$}
\Text(68.25,-34.125)[l]{$\leg{k}$}
\Text(43.6033,0.)[l]{$a$}
\Text(25.6917,13.5992)[br]{}
\Text(26.,-13.)[tr]{}
\blob
\end{picture}
}

\newcommand{\eikdiagramtwoL}{
\begin{picture}(97.5,84.5)(-22.75,-42.25)
\Line(0.,0.)(65.,32.5)
\Line(0.,0.)(65.,-32.5)
\Photon(52.,26.)(52.,-26.){1.95}{3.5}
\Photon(26.,13.)(26.,-13.){1.95}{2}
\Vertex(52.,26.){2}
\Vertex(52.,-26.){2}
\Vertex(26.,13.){2}
\Vertex(26.,-13.){2}
\Text(68.25,34.125)[l]{$\leg{j}$}
\Text(68.25,-34.125)[l]{$\leg{k}$}
\Text(58.1378,0.)[l]{$a$}
\Text(29.0689,0.)[l]{$b$}
\Text(38.5375,20.3988)[br]{}
\Text(39.,-19.5)[tr]{}
\Text(19.0271,10.6433)[br]{}
\Text(19.5,-9.75)[tr]{}
\blob
\end{picture}
}

\newcommand{\eikdiagramtwoC}{
\begin{picture}(97.5,84.5)(-22.75,-42.25)
\Line(0.,0.)(65.,32.5)
\Line(0.,0.)(65.,-32.5)
\Photon(52.,26.)(26.,-13.){-1.95}{3}
\Photon(52.,-26.)(26.,13.){-1.95}{3}
\Vertex(52.,26.){2}
\Vertex(52.,-26.){2}
\Vertex(26.,13.){2}
\Vertex(26.,-13.){2}
\Text(68.25,34.125)[l]{$\leg{j}$}
\Text(68.25,-34.125)[l]{$\leg{k}$}
\Text(46.4269,8.71038)[l]{$a$}
\Text(46.4269,-8.71038)[l]{$b$}
\Text(38.5375,20.3988)[br]{}
\Text(39.,-19.5)[tr]{}
\Text(19.0271,10.6433)[br]{}
\Text(19.5,-9.75)[tr]{}
\blob
\end{picture}
}

\newcommand{\eikdiagramtwoY}{
\begin{picture}(97.5,84.5)(-22.75,-42.25)
\Line(0.,0.)(65.,32.5)
\Line(0.,0.)(65.,-32.5)
\Photon(52.,26.)(43.6033,0.){1.95}{2}
\Photon(26.,13.)(43.6033,0.){1.95}{2}
\Photon(43.6033,0.)(43.6033,-21.8017){1.95}{2}
\Vertex(52.,26.){2}
\Vertex(26.,13.){2}
\Vertex(43.6033,0.){2}
\Vertex(43.6033,-21.8017){2}
\Text(68.25,34.125)[l]{$\leg{j}$}
\Text(68.25,-34.125)[l]{$\leg{k}$}
\Text(53.0461,12.5225)[l]{$a$}
\Text(49.5097,-11.6877)[l]{$b$}
\Text(32.2728,5.28359)[tr]{$c$}
\Text(38.5375,20.3988)[br]{}
\Text(29.25,-14.625)[tr]{}
\Text(19.0271,10.6433)[br]{}
\blob
\end{picture}
}

\newcommand{\eikdiagramtwoYinv}{
\begin{picture}(97.5,84.5)(-22.75,-42.25)
\Line(0.,0.)(65.,32.5)
\Line(0.,0.)(65.,-32.5)
\Photon(52.,-26.)(43.6033,0.){1.95}{2}
\Photon(26.,-13.)(43.6033,0.){1.95}{2}
\Photon(43.6033,0.)(43.6033,21.8017){1.95}{2}
\Vertex(52.,-26.){2}
\Vertex(26.,-13.){2}
\Vertex(43.6033,0.){2}
\Vertex(43.6033,21.8017){2}
\Text(68.25,34.125)[l]{$\leg{j}$}
\Text(68.25,-34.125)[l]{$\leg{k}$}
\Text(55.168,-13.0234)[l]{$a$}
\Text(51.6316,12.1886)[l]{$b$}
\Text(33.0387,-5.09486)[br]{$c$}
\Text(38.5375,-20.3988)[br]{}
\Text(29.25,14.625)[tr]{}
\Text(19.0271,-10.6433)[br]{}
\blob
\end{picture}
}

\newcommand{\eikdiagramthreeL}{
\begin{picture}(97.5,84.5)(-22.75,-42.25)
\Line(0.,0.)(65.,32.5)
\Line(0.,0.)(65.,-32.5)
\Line(0.,0.)(72.6722,0.)
\Photon(55.25,27.625)(55.25,0.){1.95}{2.5}
\Photon(37.05,-18.525)(37.05,0.){-1.95}{2}
\Vertex(55.25,27.625){2}
\Vertex(37.05,-18.525){2}
\Vertex(55.25,0.){2}
\Vertex(37.05,0.){2}
\Text(68.25,34.125)[l]{$\leg{k}$}
\Text(76.3058,0.)[l]{$\leg{j}$}
\Text(68.25,-34.125)[l]{$\leg{l}$}
\Text(60.4318,12.7948)[l]{$a$}
\Text(42.4369,-10.018)[l]{$b$}
\Text(46.5102,0.)[b]{}
\Text(32.1146,16.999)[br]{}
\Text(22.75,-11.375)[tr]{}
\Text(25.4353,0.)[b]{}
\blob
\end{picture}
}

\newcommand{\eikdiagramthreeLinv}{
\begin{picture}(97.5,84.5)(-22.75,-42.25)
\Line(0.,0.)(65.,32.5)
\Line(0.,0.)(65.,-32.5)
\Line(0.,0.)(72.6722,0.)
\Photon(37.05,18.525)(37.05,0.){1.95}{2}
\Photon(55.25,-27.625)(55.25,0.){-1.95}{2.5}
\Vertex(37.05,18.525){2}
\Vertex(55.25,-27.625){2}
\Vertex(37.05,0.){2}
\Vertex(55.25,0.){2}
\Text(68.25,34.125)[l]{$\leg{k}$}
\Text(76.3058,0.)[l]{$\leg{j}$}
\Text(68.25,-34.125)[l]{$\leg{l}$}
\Text(60.1189,-14.1922)[l]{$a$}
\Text(42.4369,10.018)[l]{$b$}
\Text(46.5102,0.)[t]{}
\Text(32.1146,16.999)[br]{}
\Text(22.75,-11.375)[tr]{}
\Text(25.4353,0.)[t]{}
\blob
\end{picture}
}

\newcommand{\eikdiagramthreeY}{
\begin{picture}(97.5,84.5)(-22.75,-42.25)
\Line(0.,0.)(65.,32.5)
\Line(0.,0.)(65.,-32.5)
\Line(0.,0.)(72.36,-6.72921)
\Photon(45.5,22.75)(50.3792,7.05301){1.95}{2}
\Photon(39.,-19.5)(50.3792,7.05301){-1.95}{3}
\Photon(62.9532,-5.85441)(50.3792,7.05301){1.95}{2}
\Vertex(45.5,22.75){2}
\Vertex(62.9532,-5.85441){2}
\Vertex(39.,-19.5){2}
\Vertex(50.3792,7.05301){2}
\Text(68.25,34.125)[l]{$\leg{j}$}
\Text(75.978,-7.06567)[l]{$\leg{k}$}
\Text(68.25,-34.125)[l]{$\leg{l}$}
\Text(52.0477,16.1784)[l]{$a$}
\Text(61.6758,3.43505)[l]{$b$}
\Text(45.4209,-12.9719)[l]{$c$}
\Text(26.,-13.)[tr]{}
\Text(28.9031,15.2991)[br]{}
\Text(29.0689,0.)[b]{}
\blob
\end{picture}
}

\newcommand{\eikdiagramfourL}{
\begin{picture}(97.5,84.5)(-22.75,-42.25)
\Line(0.,0.)(65.,32.5)
\Line(0.,0.)(65.,-32.5)
\Line(0.,0.)(72.36,6.72921)
\Line(0.,0.)(72.36,-6.72921)
\Photon(52.,26.)(57.888,5.38336){1.95}{2}
\Photon(52.,-26.)(57.888,-5.38336){-1.95}{2}
\Vertex(52.,26.){2}
\Vertex(52.,-26.){2}
\Vertex(57.888,5.38336){2}
\Vertex(57.888,-5.38336){2}
\Text(68.25,34.125)[l]{$\leg{j}$}
\Text(75.978,7.06567)[l]{$\leg{k}$}
\Text(75.978,-7.06567)[l]{$\leg{l}$}
\Text(68.25,-34.125)[l]{$\leg{m}$}
\Text(59.3965,16.9633)[l]{$a$}
\Text(59.3965,-16.9633)[l]{$b$}
\blob
\end{picture}
}


\thispagestyle{empty}
\def\thefootnote{\fnsymbol{footnote}}
\setcounter{footnote}{1}
\null
\draftdate\hfill  PSI-PR-03-01 \\
\strut\hfill  TTP-03-03\\
\strut\hfill hep-ph/0301241
\vskip 0cm
\vfill
\begin{center}
{\Large \bf
Two-loop electroweak angular-dependent logarithms at high energies
\par}
 \vskip 1em
{\large
{\sc A.\ Denner$^1$\footnote{Ansgar.Denner@psi.ch}, 
M.\ Melles$^1$\footnote{Michael.Melles@psi.ch},
and S.\ Pozzorini$^2$\footnote{pozzorin@particle.uni-karlsruhe.de} }}
\\[.5cm]
$^1$ {\it Paul Scherrer Institut\\
CH-5232 Villigen PSI, Switzerland}
\\[0.3cm]
$^2$ {\it Institut f\"ur Theoretische Teilchenphysik, 
Universit\"at Karlsruhe \\
D-76128 Karlsruhe, Germany}
\par
\end{center}\par
\vskip 1.0cm 
\vfill 
{\bf Abstract:} \par 

We present results on the two-loop leading and angular-dependent
next-to-leading logarithmic virtual corrections to arbitrary processes
at energies above the electroweak scale. In the `t~Hooft--Feynman
gauge the relevant Feynman diagrams involving soft and collinear gauge
bosons $\gamma, \PZ, \PW^\pm$ coupling to external legs are evaluated
in the eikonal approximation in the region where all kinematical
invariants are much larger than the electroweak scale.  The
logarithmic mass singularities are extracted from massive multi-scale
loop integrals using the Sudakov method and alternatively the
sector-decomposition method in the Feynman-parameter representation.
The derivations are performed within the spontaneously broken phase of
the electroweak theory, and the two-loop results are in agreement with
the exponentiation prescriptions that have been proposed in the
literature based on a symmetric $\SUtwo\times\Uone$ theory matched
with QED at the electroweak scale.

\par
\vskip 1cm
\noindent
January 2003 
\par
\null
\setcounter{page}{0}
\clearpage
\def\thefootnote{\arabic{footnote}}
\setcounter{footnote}{0}

\section{Introduction}

The main task of future colliders such as the LHC
\cite{Haywood:1999qg} or an $\Pep\Pem$ Linear Collider (LC)
\cite{Accomando:1998wt,Aguilar-Saavedra:2001rg,Abe:2001wn,Abe:2001gc} 
will be the
investigation of the origin of electroweak symmetry breaking and the
exploration of the limits of the Electroweak Standard Model.  In order
to disentangle effects of physics beyond the Standard Model, the
inclusion of QCD and electroweak radiative corrections into the
theoretical predictions is crucial.

In the energy range of future colliders, \ie at energies above the
electroweak scale, $\sqrt{s}\gg M\simeq\MW\simeq\MZ$, the electroweak
corrections are enhanced by large logarithmic contributions
\cite{Kuroda:1991wn} of the type
\beq\label{logform}
\alpha^L\log^{N}{\left(\frac{s}{M^2}\right)},\quad 1\le N \le 2L.
\eeq
These electroweak logarithmic corrections (EWLC) can be classified in
a gauge-invariant way according to the powers $N$ of the logarithms of
$s/M^2$.  The leading logarithms (LL), also known as Sudakov
logarithms \cite{Sudakov:1954sw}, correspond to $N=2L$, the
next-to-leading logarithms (NLL) to $N=2L-1$, etc.

The above logarithmic terms constitute the singular part of the
corrections in the massless limit, $M^2/s\to 0$.  They result either
as remnants of ultraviolet singularities after parameter
renormalization, or as mass singularities from soft/collinear emission
of virtual or real particles off initial or final-state particles.
These latter do not cancel in observables, in contrast to the
well-known soft and collinear singularities observed in QCD. This is,
on the one hand, due to the fact that the masses of the weak gauge
bosons provide a physical cut-off, and that there is no need to
include real Z-boson and W-boson bremsstrahlung.  On the other hand,
the Bloch--Nordsieck theorem is violated also in inclusive quantities
in non-abelian gauge theories if the asymptotic states carry
non-abelian charges or in spontaneously broken abelian gauge theories
if mass eigenstates result from mixing of gauge eigenstates
\cite{Ciafaloni:2000df}. This leads to the appearance of LL also in
inclusive quantities in such theories and thus in the electroweak
Standard Model.
As a consequence, the electroweak corrections can become of the order
of the QCD corrections in the TeV energy range.

These enhanced EWLC have found quite some interest recently; for
reviews we refer to \citeres{Melles:2001ye,Denner:2001mn}.  At the
one-loop level the EWLC have been obtained, on the one hand, via
explicit diagrammatic calculations for many $2\to2$ scattering
processes in the Standard Model and the Minimal Supersymmetric
Standard Model
\cite{Beenakker:1993tt,Ciafaloni:1999xg,Beccaria:2000fk,Beccaria:2001yf,Layssac:2001ur}.
On the other hand, the universality of the electroweak LL and NLL has
been proven, and general results have been given for arbitrary
electroweak processes that are not mass-suppressed at high energies
\cite{Denner:2001jv,Pozzorini:rs} and applied, for instance, 
to gauge-boson pair production at the LHC \cite{Accomando:2001fn}.

The typical size of the one-loop electroweak corrections from LL and NLL 
for a  $2\to 2$  cross section is%
\beq\label{oneloopestimate}
-\frac{\alpha}{\pi\sw^2}\log^2\left(\frac{s}{M^2}\right)\simeq -26\%
,\qquad
+\frac{3\alpha}{\pi\sw^2}\log\left(\frac{s}{M^2}\right)\simeq 16\%,
\eeq
respectively, at $\sqrt{s}=1\TeV$, with $M=\MW$, and
$1-\sw^2=\cw^2=\MW^2/\MZ^2$.  The size of the corrections increases
with the number of particles in the final state, and it is important
to note that at the TeV scale, the LL and NLL have similar size and
opposite sign resulting in large cancellations
\cite{Beccaria:2001yf,Kuhn:2000nn}.

Resummations of the EWLC have been proposed based on techniques and
results known from QCD and QED.  Fadin \etal\ \cite{Fadin:2000bq} have
resummed the LL by means of the infrared evolution equation (IREE),
which describes the all-order leading-logarithmic dependence of a
matrix element with respect to the transverse-momentum cut-off
$\mu_\perp$, within a symmetric gauge theory.  This equation was
applied to the electroweak theory by assuming that the
$\mu_\perp$-integration can be split into two regimes both
corresponding to symmetric gauge theories.  In the regime $s\ge
\mu_\perp\ge M$, $\SUtwo\times\Uone$ symmetry was used, whereas for
$M\ge \mu_\perp\ge \la$ the $\Uone_\elm$ symmetry was assumed.
K\"uhn \etal\ have applied results from QCD to resum the logarithmic
corrections to massless 4-fermion processes, $\Pep\Pem\to\Pf\bar\Pf$,
up to the NLL \cite{Kuhn:2000nn} and even to the
next-to-next-to-leading logarithms (NNLL) \cite{Kuhn:2001hz}.
This was done for a symmetric $\SUtwo\times\Uone$ theory 
and additional electromagnetic effects were included following the
IREE approach.  It was found that at $1\TeV$ there is no clear
hierarchy between LL, NLL, and NNLL.  One of us has proposed a
resummation of the NLL for arbitrary processes
\cite{Melles:2001gw,Melles:2001dh}, which relies on the prescription
of matching a symmetric $\SUtwo\times\Uone$ theory with QED at the
electroweak scale.

All these resummations amount to exponentiations of the EWLC.  The
approximate size of the two-loop LL and NLL resulting from the
exponentiation of the one-loop corrections \refeq{oneloopestimate} for
$2\to 2$ processes at $\sqrt{s}=1\TeV$ is
\beq\label{twoloopestimate}
+\frac{\alpha^2}{2\pi^2\sw^4}\log^4\left(\frac{s}{M^2}\right)\simeq 3.5\%
,\qquad
-\frac{3\alpha^2}{\pi^2\sw^4}\log^3\left(\frac{s}{M^2}\right)\simeq -4.2\%
,
\eeq
respectively, and it is clear that in view of the precision objectives
of a LC below the per-cent level these two-loop EWLC must be under
control.

All the above resummation prescriptions result from matching a {\em
  symmetric} $\SUtwo\times\Uone$ theory and QED at the electroweak
scale, and are based on the assumption that other effects related to
{\em spontaneous symmetry breaking} may be neglected at high energies.
This assumption needs to be checked by explicit diagrammatic two-loop
calculations based on the electroweak Lagrangian, where all relevant
effects related to spontaneous symmetry breaking are taken into
account.  In particular, the following non-trivial aspects need to be
treated with care: (i) There is a multi-scale hierarchy of masses,
$M\gg \Mfl\gg \la$, with heavy masses $\Mt\sim \MH\sim M$ at the
electroweak scale, light-fermion masses $\Mfl$, and also an
infinitesimal photon mass $\la$, which is used as infrared regulator.
As a consequence of this hierarchy, also logarithms of the large
ratios ${M}/\Mfl$ and ${\Mfl}/\la$ have to be taken into account, and the
general form of logarithmic terms of order $N$ in \refeq{logform}
becomes
$\log^{N_1}(s/M^2)\log^{N_2}(M^2/\Mfl^2)\log^{N_3}(\Mfl^2/\la^2)$,
with $N=N_1+N_2+N_3$.  (ii) In the gauge-boson sector, the gauge-group
eigenstates $B,W^3$ mix resulting in mass eigenstates $\gamma,\PZ$
with a large mass gap $\la \ll M$.  (iii) Longitudinal gauge bosons
appear as physical asymptotic states.

The resummation of the LL has been checked for the massless fermionic
singlet form factor in \citeres{Melles:2000ed,Hori:2000tm} and for
arbitrary processes in the Coulomb gauge in \citere{Beenakker:2000kb}.
The resummation of the next-to-leading logarithms has so far not been
confirmed by explicit two-loop calculations.  In the present paper, we
consider a specific gauge-invariant subset of the next-to-leading EWLC
to exclusive processes: the angular-dependent NLL of type
\beq\label{anglogform}
\alpha^L\log^{2L-1}{\left(\frac{s}{M^2}\right)}
\log{\left(\frac{|r|}{s}\right)},
\eeq
where $r$ represents a generic kinematic invariant different from $s$,
and the ratio $r/s$ depends on the angles between external momenta.
These angular-dependent NLL are numerically important as has been
stressed in \citeres{Beccaria:2001yf,Kuhn:2000nn,Kuhn:2001hz}.
Prescriptions for their resummation have been given in
\citeres{Kuhn:2000nn,Kuhn:2001hz} for massless 4-fermion processes and
extended to arbitrary processes in \citere{Melles:2001dh}.  As we have
mentioned above, these prescriptions are based on symmetric gauge
theories.  The purpose of this paper is to check them with an explicit
two-loop calculation within the spontaneously broken electroweak
theory.

At one-loop order, in the 't~Hooft--Feynman gauge, the
angular-dependent NLL result only from diagrams where a gauge boson is
exchanged between two external lines.  Similarly, the
angular-dependent NLL at two-loop order can be traced back to a
relatively small set of Feynman diagrams. This allows us to present a
diagrammatic calculation of the two-loop angular-dependent NLL for
arbitrary processes. The calculation is based on the eikonal
approximation. The relevant massive two-loop integrals are evaluated
analytically using two independent methods, one goes back to Sudakov
\cite{Sudakov:1954sw}, the other uses sector decomposition of
Feynman-parameter integrals
\cite{Hepp:1966eg,Roth:1996pd,Binoth:2000ps}.

The paper is organized as follows: In \refse{se:heapp0} we define our
conventions and the approximations used in the high-energy limit and
we discuss the Feynman diagrams that give rise to the leading mass
singularities and the eikonal approximation.  Section~\ref{se:loopint}
is devoted to the description of the calculation of the two-loop
integrals. The results for the contributing diagrams and their sum are
presented in \refse{se:results}.  The appendices provide information
on our conventions and the explicit results of the individual loop
integrals as well as relations between them.

\section{High-energy logarithmic approximation}
\label{se:heapp0}
\subsection{Preliminaries}
\label{se:heapp}
We consider generic electroweak processes involving $n$ arbitrary
polarized particles, which may be light or heavy chiral fermions,
transverse or longitudinal gauge bosons, or Higgs bosons.  As a
convention, we consider $n\to 0$ processes,
\beq\label{process}
\varphi_{i_1}(p_1)\ldots \varphi_{i_n}(p_n)\to 0,
\eeq
where all particles $\varphi_{i_k}$ and their momenta $p_k$ are
assumed to be incoming. Corresponding $2\to n-2$ processes are easily
obtained by crossing symmetry.  Our calculations are performed in the
physical basis, where the external particles $\varphi_{i_k}$ as well
as the virtual particles in the loops correspond to mass eigenstates,
and mixing effects are properly taken into account.  The matrix
elements for the processes \refeq{process} and the external-leg gauge
couplings are denoted with the shorthands
\beq\label{mel}
\M^{{i_1}\ldots {i_n}}
\equiv\M^{\varphi_{i_1}\ldots \varphi_{i_n}}(p_1,\dots,p_n),
\qquad
I^a_{i'_ki_k}\equiv I^{V^a}_{\varphi_{i'_k}\varphi_{i_k}},
\eeq
where $I^a_{i'i}$ corresponds to the coupling of the gauge bosons
$V^a=\gamma,\PZ,\PWpm$
to an incoming particle 
$\varphi_{i}$ 
and an outgoing particle 
$\varphi_{i'}$.  More details concerning gauge couplings can be found
in \refapp{se:couplings}.

All external-leg momenta are assumed to be on shell, $p_k^2=m_k^2$,
and we restrict ourselves to the high-energy region where
$s=(p_1+p_2)^2\approx2p_1p_2$ as well as all other kinematical
invariants are much larger than the electroweak scale.  In particular,
we assume the following hierarchy of energy and mass
scales%
\footnote{The first inequality implies, in particular, that all angles
  are larger than $M/\sqrt{s}$ in reference frames where all particle
  energies are not much larger than $\sqrt{s}$.}
\beq\label{masshierarchy}
|(p_k+p_l)^2|\simeq |2p_kp_l| 
\gg M^2 \simeq\MZ^2\simeq \MW^2
\gg m_{f\ne\Pt}^2 
\gg M_\gamma^2\equiv\la^2.
\eeq
The mass scale $M$ is used to denote a generic weak-boson mass in the
logarithms, and we neglect logarithms of the ratio $\MW/\MZ$, which
originate from the difference between the \PZ- and the \PW-boson mass.
With $\Mfl$ we denote the masses of the light fermions.  The
infinitesimal photon mass $\la$ is used to regularize infrared
divergences.

All masses of real and virtual particles are assumed to be at or below the
electroweak scale.
Nevertheless, our results also apply to processes involving particles
with masses
that are heavier but of the same order as $M$, \ie light Higgs bosons
or top quarks with $\MH\simeq\Mt\simeq M$.  However, the logarithms
involving the ratios $\MH/M$ and $\Mt/M$
are neglected.

Our next-to-leading logarithmic angular-dependent (\NLLad)
approximation is defined as follows.  We consider corrections that are
logarithmically divergent in the limit where the ratios $M/\sqrt{s}$,
$\Mfl/M$, and $\la/\Mfl$ vanish.  From these mass-singular logarithms
we retain only the leading and the next-to-leading angular-dependent
ones, \ie at $L$ loops (with $L=1,2$) we consider only contributions
of the order
\beqar\label{LA}
&&\ord\left[
\alpha^L
\log^{2L-N}\left(\frac{s}{M^2}\right)
\prod_{i=1}^N\log\left(\frac{M^2}{m_{\mathrm{light},i}^2}\right)
\right],\quad  0\le N \le 2 L,\quad\mbox{and} 
\nl
&&\ord\left[
\alpha^L
\log\left(\frac{|2p_{k}p_{l}|}{s}\right)
\log^{2L-N-1}\left(\frac{s}{M^2}\right)
\prod_{i=1}^N\log\left(\frac{M^2}{m_{\mathrm{light},i}^2}\right)
\right],\quad  0\le N \le 2 L -1,
\eeqar
where $m_{\mathrm{light},i}$ are either light-fermion or photon
masses.  As stated above, the logarithms of ratios of heavy masses are
neglected, \ie we consider $\log{\MZ}\simeq \log{\MW} \simeq \log{\MH}
\simeq \log{\Mt}$.
 
In our approximation all terms that are suppressed by factors
$M/\sqrt{s}$, $\Mfl/M$ or $\la/\Mfl$ are neglected. In practice, all
mass terms in the numerators of loop integrals are omitted in the
calculations.  In order to avoid factors of order $\sqrt{s}/M$ from
longitudinal polarization vectors that would enhance mass-suppressed
contributions, the Goldstone-Boson Equivalence Theorem (GBET)
\cite{et} has to be used for matrix elements involving longitudinal
gauge bosons.  For our purposes we can use the GBET in its naive
lowest-order form since the quantum corrections to the GBET involve
only two-point functions, which give no contribution in the considered
\NLLad\ order \refeq{LA} in the 't~Hooft--Feynman gauge.  In practice
each longitudinal gauge boson $V^a_\rL=\PW^\pm_\rL,\PZ_\rL$ has to be
substituted by a corresponding would-be Goldstone boson
$\Phi_a=\phi^\pm,\chi$ using
\beq\label{gbet}
\M^{\varphi_{i_1}\ldots V^a_{\rL}\ldots \varphi_{i_n}}=
\ri^{(1-Q_{V^a})} \M^{\varphi_{i_1}\ldots \Phi_a\ldots \varphi_{i_n}},
\eeq
where $Q_{V^a}=\pm 1,0$ is the corresponding charge.  Thus, the
general results for EWLC presented in the following have to be applied
to the matrix elements involving would-be-Goldstone bosons.

We restrict ourselves to matrix elements that are not mass-suppressed
in the high-energy limit. This permits us to make use of the
identity%
\footnote{Here the sums over the components $i'_k$ of the multiplets
  corresponding to the various external particles $i_k$ are implicitly
  understood.}
\beq\label{eq:gc}
\sum_{k=1}^n  \M^{{i_1}\ldots {i'_k} \ldots {i_n}}\, I^a_{i'_ki_k} = 
\ord\left(\frac{M}{\sqrt{s}}
\right)\,\M^{{i_1}\ldots {i_n}}\simeq 0,
\eeq
which can be derived from global $\SU(2)\times\U(1)$ symmetry and is
very useful in order to simplify sums over external-leg insertions of
the gauge-group generators.

\subsection{Feynman diagrams in eikonal approximation}
\label{se:eikonal}

Mass singularities originate from diagrams with virtual particles
coupling to on-shell external legs.  In this paper we consider only
$\NLLad$ contributions that result from the leading mass
singularities, \ie at two loops from contributions involving four
mass-singular logarithms.  We perform the calculation in the
't~Hooft--Feynman gauge, where the gauge-boson propagators have the
same pole structure as scalar propagators and the leading mass
singularities originate only from diagrams with soft--collinear
virtual gauge bosons coupling to different external particles. The
exchange of soft--collinear scalar particles or fermions is
mass-suppressed.  The relevant two-loop diagrams have the structure
\beqar\label{diagrams}
\diagtwoL{jk}{ab}&=&
\vcenter{\hbox{\smalldiagramtwoL}}
,\quad
\diagtwoC{jk}{ab}=
\vcenter{\hbox{\smalldiagramtwoC}}
,\quad
\diagtwoY{jk}{abc}=
\vcenter{\hbox{\smalldiagramtwoY}},\nl
\diagthreeL{jkl}{ab}&=&
\vcenter{\hbox{\smalldiagramthreeL}}
\hspace{2mm},\quad
\diagthreeY{jkl}{abc}=
\vcenter{\hbox{\smalldiagramthreeY}}
\hspace{2mm},\quad
\diagfourL{jklm}{ab}=
\vcenter{\hbox{\smalldiagramfourL}}\hspace{2mm},\;\nln
\eeqar
where the  soft--collinear gauge bosons $V^a,V^b,V^c=\gamma,\PZ,\PWpm$
are  exchanged between two, 
three, 
or four 
of the $n$ on-shell external legs $j,k,l,m=1,\dots,n$.  The external
lines, which are generically represented by full lines, can be
fermions, transverse gauge bosons, would-be Goldstone bosons, or Higgs
bosons.  The external legs that do not couple to the soft gauge bosons
are represented by the lines and the dots on the left-hand side of the
grey blobs.

In order to extract the leading mass singularities, the
Feynman diagrams \refeq{diagrams} are evaluated in eikonal
approximation, \ie by approximating the integrand in the limit where
the momenta of the soft gauge bosons are small.  In this limit, the
above Feynman diagrams can be treated
independently of the process and the spin
of the external particles, \ie universally for 
chiral fermions, transverse gauge bosons, or scalar particles. This is
done as follows:
\begin{itemize}
\item In the ``hard part'' of the diagrams corresponding to the grey
  blobs in \refeq{diagrams} the momenta of the soft gauge bosons are
  neglected, and only the remaining ``soft part'' of the diagrams has
  to be integrated over the loop momenta.  The blobs typically involve
  contributions from various tree diagrams, but they do not need to be
  evaluated explicitly.  In our derivations we only make use of the
  charge-conservation identity \refeq{eq:gc} to relate the complete
  tree-level amplitudes from different blobs.

\item The vertices involving three soft gauge bosons are associated
  with the usual Yang--Mills couplings
\beqar
\begin{array}{l}
\vcenter{\hbox{
\begin{picture}(110,90)(-50,-45)
\Text(-50,5)[lb]{\scriptsize$V^{a_1}_{\mu_1}(l_1)$}
\Text(35,30)[b]{\scriptsize$V^{a_2}_{\mu_2}(l_2)$}
\Text(35,-30)[t]{\scriptsize$V^{a_3}_{\mu_3}(l_3)$}
\Vertex(0,0){2}
\Photon(0,0)(35,25){2}{3.5}
\Photon(0,0)(35,-25){2}{3.5}
\Photon(0,0)(-45,0){2}{3.5}
\end{picture}}}
\end{array}
&&
\begin{array}{l}
\\
=
-\ri e I^{{a_1}}_{\bar{a}_3\,{a_2}}
\Bigl[
g_{\mu_1\mu_2}(l_1-l_2)_{\mu_3}
+g_{\mu_2\mu_3}(l_2-l_3)_{\mu_1}
\\
\hphantom{=\frac{e}{\sw}\varepsilon^{V^{a_1}V^{a_2}V^{a_3}}\Bigl[}
+g_{\mu_3\mu_1}(l_3-l_1)_{\mu_2}\Bigr],
\end{array}
\eeqar
where the particles and momenta are incoming,
$I^{{a_1}}_{\bar{a}_3\,{a_2}}$ are the structure constants defined by
\refeq{structurefun}, and $\bar{a}_i$ indicates the complex conjugate
of $a_i$.  In the 't~Hooft--Feynman gauge, the propagators read $-\ri
g^{\mu\nu}/(l^2-M_a^2+\ri\varepsilon)$ for a gauge boson $V^a$.

\item
The emission of gauge bosons with momenta $l_1,l_2,\ldots$
along an incoming external line with momentum $k_1=p_\mathrm{ext}$
in the soft limit  $l_i\to 0$, 
\beq\label{eikvertexc}
\vcenter{\hbox{\begin{picture}(90,60)(0,-30)
\Line(90,0)(0,0)
\LongArrow(10,5)(20,5)
\LongArrow(40,5)(50,5)
\LongArrow(70,5)(80,5)
\LongArrow(25,-10)(25,-20)
\LongArrow(55,-10)(55,-20)
\Line(90,0)(0,0)
\Photon(30,0)(30,-30){2}{3}
\Vertex(30,0){2}
\Photon(60,0)(60,-30){2}{3}
\Vertex(60,0){2}
\Text(15,10)[b]{\scriptsize$k_1$}
\Text(45,10)[b]{\scriptsize$k_2$}
\Text(75,10)[b]{\scriptsize$k_3$}
\Text(20,-15)[r]{\scriptsize$l_1$}
\Text(50,-15)[r]{\scriptsize$l_2$}
\Text(92,0)[l]{\scriptsize{$\dots$}}
\end{picture}}}
\quad\,,
\eeq
gives rise to a product of terms containing a factor
$\ri/(k_j^2-m_j^2+\ri\varepsilon)$ for each propagator with momentum
$k_j=k_{j-1}-l_{j-1}$ and mass $m_j$ and an eikonal factor
\cite{Pozzorini:rs}
\beq\label{eikvertexb}
\vcenter{\hbox{\begin{picture}(80,60)(5,-30)
\Line(70,0)(-10,0)
\Photon(30,0)(30,-25){2}{3}
\Vertex(30,0){2}
\Text(25,-20)[r]{\scriptsize$\bar{V}^a_\mu(-l_j)$}
\Text(-10,5)[lb]{\scriptsize$\varphi_{i}(k_j)$}
\Text(35,5)[lb]{\scriptsize$\bar{\varphi}_{i'}(-k_{j+1})$}
\end{picture}}}
\EIK
2k_j^\mu\, \ri eI^{\bar a}_{i' i},   
\eeq
for each vertex, where $I^a_{i'i}$ are the generators defined in
\refapp{se:couplings}.  Note that the form of these eikonal factors
only depends on the gauge-group representation of the inflowing
particles, but not on their spin. With $\EIK$ we denote equations that
are valid within the eikonal approximation.

\item The eikonal factors defined above are proportional to the
  momenta of the emitting particles, and if these particles are
  virtual they depend on the loop momentum $l_{1}$ via
  $k_{2}=p_{\mathrm{ext}}-l_1$.  If the integration is restricted to
  the region of soft gauge-boson momenta, $l_{1}\approx 0$, as in the
  Sudakov method, this loop-momentum dependence can be neglected, and
  one can use $k_{2}=p_\mathrm{ext}$ in the eikonal factors.  However,
  if one integrates over the full loop-momentum space, as in the
  Feynman-parameter representation, also the region $k_2=
  p_{\mathrm{ext}}-l_{1}\approx 0$ where the emitting line becomes
  soft may be important.  This happens for the diagrams
  $\symbdiag_{\mathrm{2C}}$ and $\symbdiag_{\mathrm{3L}}$, where such
  regions give rise to spurious leading logarithms if one uses eikonal
  factors with $k_{2}=p_\mathrm{ext}$, whereas the
  loop-momentum-dependent eikonal factors defined in
  \refeq{eikvertexb} suppress them\footnote{For all other topologies
    the loop-momentum dependence of the eikonal factors is irrelevant
    in \NLLad\ approximation.}.  The reason why these contributions
  have to be suppressed is the following: if the emitting line in
  \refeq{eikvertexc} is a fermion or a scalar particle the mass
  singularity for $k_{2}=0$ gets mass-suppressed in the complete
  Feynman diagram by the numerator.  If the emitting line is a
  transverse gauge boson this remains true for the region $k_2\approx
  l_2\approx0$, whereas the region with $k_2\approx 0$ and $l_2\approx
  p'_\mathrm{ext}\neq p_\mathrm{ext}$, which gives leading
  contributions only to the diagram $\symbdiag_{\mathrm{2C}}$, has to
  be suppressed in order to avoid double counting of topologically
  equivalent configurations when the sum over all soft--collinear
  gauge bosons is performed.
\item The denominators of the propagators denoted by full lines in
  \refeq{diagrams} are kept exact, \ie the square of the momenta of
  the soft gauge bosons is not neglected there.
\end{itemize}
The explicit expressions for the diagrams \refeq{diagrams} in eikonal
approximation are given in \refapp{se:twoloopint}.

All terms involving four mass-singular logarithms originate from the
diagrams \refeq{diagrams} and there only from the terms that are kept
in the eikonal approximation. Other contributions give rise to at most
three mass-singular logarithms.  In this paper we assume that all
angular-dependent NLL result only from contributions with four
mass-singular logarithms via the appearance of different scales in the
logarithms. This is equivalent to the assumption that generic NLL are
not multiplied by logarithms of ratios of kinematical variables that
do not result from mass singularities.  Although we have not proven
this assumption so far, we do not see a source for additional
angular-dependent NLL. We have checked for several examples that no
such terms arise from neglected contributions to the diagrams
\refeq{diagrams}, \ie terms with gauge-boson momenta in the
numerators. A complete proof of the assumption, however, requires a
calculation of the complete set of NLL.

\subsection{Mass-gap effects}
\label{se:massgap}

Each topology in \refeq{diagrams} has to be evaluated for all
different mass assignments corresponding to the various electroweak
gauge bosons $V^a,V^b,V^c=\gamma,\PZ,\PWpm$.  In practice, for each
diagram involving two or three soft--collinear gauge bosons we have
the following four cases
\beqar\label{smcases}
(M_a,M_b)&=&(\la,\la), (\la,M), (M,\la), (M,M),
\label{smcase2}
\\
(M_a,M_b,M_c)&=&(M,M,M), (\la,M,M), (M,\la,M), (M,M,\la),
\label{smcase3}
\eeqar
whereas the external lines are assumed to have arbitrary masses
$m_k^2$ at or below the electroweak scale.

Our main aim is to investigate the effects related to symmetry
breaking, and in particular the effects of the large mass gap $\la\ll
M$ in the gauge sector.  This gives rise to logarithms of the photon
mass $\la$ and light-fermion masses that violate $\SUtwo\times\Uone$
symmetry, such that the full electroweak result is only symmetric with
respect to the unbroken $\Uone_{\elm}$ group.  Nevertheless, according
to the physical picture proposed in the IREE approach
\cite{Fadin:2000bq} and generalized by other authors
\cite{Kuhn:2000nn,Kuhn:2001hz,Melles:2001gw,Melles:2001dh}, the EWLC
are expected to exhibit a higher degree of symmetry.  In this picture,
the complete electroweak result factorizes into a part which exhibits
$\SUtwo\times\Uone$ symmetry and corresponds to the full electroweak
result for the case $\la=M$, and a remaining part that originates from
the mass gap $\la\ll M$ and exhibits $\Uone_{\elm}$ symmetry.

In order to check this picture at the level of angular-dependent NLL,
we organize our calculation as follows.  All intermediate results
$f(\la)$ depending on the photon mass $\la$ are split into two parts
as
\beq\label{eq:subtraction}
f (\la)\equiv f (M)+\Delta f (\la),
\eeq
where the part $f (M)$ corresponds to the case $\la=M$ and has to be
calculated for%
\footnote{Note that this part $f(M)$ cannot be obtained by simply
  substituting $\la=M$ in the result $f(\la)$ since the inequalities
  \refeq{masshierarchy} and \refeq{massgap2} exclude each
  other.}
\beq\label{massgap2}
\la=M \gg \Mfl.
\eeq
In this case, all mass singularities are regulated by $M$ and the
light fermion masses below the electroweak scale $\Mfl\ll M$ can be
neglected.  The remaining part, $\Delta f(\la)=f(\la)-f(M)$,
originates from the mass gap $\la\ll M$.  In the language of the IREE,
this subtracted part can be understood as the part of the photonic
contribution that originates from below the electroweak scale.

\subsection{Validity of our results for  extensions of the electroweak
  theory} 

Our derivations depend only on a few general features of the
Electroweak Standard Model, such as the underlying global gauge
symmetry, the spectrum of gauge bosons, and the fact that all particle
masses are of the order of the electroweak scale or lighter. All
leading and next-to-leading angular-dependent logarithms originate
only from (soft-collinear) gauge bosons and depend only on gauge
couplings.  Therefore our results apply also to those extensions of
the Electroweak Standard Model, where these features are preserved,
\ie where no additional gauge bosons and no logarithms involving mass
scales much higher than the electroweak scale appear.

Such models include softly broken supersymmetric extensions such as
the Minimal Supersymmetric Standard Model in the case where the masses
of the superpartner particles are of the order of the electroweak
scale $M$. Owing to mixing, the gauge couplings $I^a_{i'i}$ may
involve mixing matrices.  More details about higher-order
supersymmetric EWLC can be found in \citere{Beccaria:2001fu}.

\section{Loop integrals in logarithmic approximation}
\label{se:loopint}
The loop integrals were evaluated in logarithmic approximation using
two independent methods: the Sudakov technique and the
sector-decomposition method described in \refse{se:sudakov} and
\refse{se:sectordecomp}, respectively.  These two methods were applied
in a complementary way in order to calculate all integrals in
\refapp{app:integrals} and to perform various cross checks.
  
\subsection{The Sudakov technique for angular-dependent logarithms} 
\label{se:sudakov}

\newcommand{\bl}{{\bf l}}

The Sudakov technique has long been known and used for the calculation
of the leading logarithmic asymptotics of field theories
\cite{Sudakov:1954sw,lan}.  In this paper, we apply this approach to
the calculation of the leading and next-to-leading angular-dependent
logarithms at the two-loop level. To our knowledge, the Sudakov
technique has so far only been applied to ladder or crossed ladder
diagrams involving a single large invariant.  Here we generalize this
method to the case of different large invariants.

We illustrate the method for the 3-leg ladder diagram
$\diagthreeL{jkl}{ab}$ shown in \refeq{3legsladder}. In the eikonal
approximation the corresponding integral is given by \refeq{eq:s23}.
It involves four different mass singularities and thus gives rise to
four large logarithms.  These singularities appear if the gauge-boson
momenta $l_1$ and $l_2$ become soft and collinear to the momenta of
the external particles to which the gauge bosons couple. Mass terms
have to be only retained as far as they regularize the mass
singularities.

In order to extract these singularities it is convenient to use the
following Sudakov parametrizations for the loop momenta,
\beqar
l_1&=&y_1 \left(p_j-\frac{m_j^2}{2p_jp_k} p_k\right)
   +x_1\left( p_k-\frac{m_k^2}{2p_jp_k} p_j \right)+l_{1,\perp },\nl
l_2&=&y_2 \left(p_j-\frac{m_j^2}{2p_jp_l} p_l\right)
   +x_2\left( p_l-\frac{m_l^2}{2p_jp_l} p_j \right)+l_{2,\perp },
\label{eq:lpdefm}
\eeqar
where $l_{1,\perp }p_j=l_{1,\perp }p_k=l_{2,\perp}p_j =
l_{2,\perp}p_l=0$.  The mass terms turn out to be
only relevant for photon exchange since we assume $m_{j,k,l} \lsim M$.
The two-dimensional transverse momenta $l_{i,\perp }$ are space-like.
They can be parametrized by their moduli $|\bl_{i,\perp}|$ and
azimuthal angles $\phi_i$.  Then up to irrelevant mass terms, the
integration measures read $\rd^4l_1=\frac{1}{2} |p_j
p_k|\rd\bl^2_{1,\perp} \rd \phi_1 \rd x_1 \rd y_1$ and
$\rd^4l_2=\frac{1}{2} |p_j p_l|\rd \bl^2_{2,\perp} \rd \phi_2 \rd x_2
\rd y_2$.

Since the leading logarithms originate from the regime of soft and
collinear gauge-boson momenta, we can drop all $l$-dependent terms in
the numerator and replace the gauge-boson propagators as
\beqar \label{eq:propid}
\frac{\ri}{l_1^2- M_a^2+\ri\varepsilon}&=&
\frac{\ri}{2p_jp_k  x_1 y_1 - M_a^2-\bl^2_{1,\perp}+\ri\varepsilon} 
\to
\pi \delta ( 2p_jp_k  x_1 y_1 - M_a^2-\bl^2_{1,\perp}),\nl
\frac{\ri}{l_2^2- M_b^2+\ri\varepsilon}&=&
\frac{\ri}{2p_jp_l  x_2 y_2 - M_b^2-\bl^2_{2,\perp}+\ri\varepsilon} 
\to
\pi \delta ( 2p_jp_l  x_2 y_2 - M_b^2-\bl^2_{2,\perp}),
\eeqar
up to irrelevant terms of order $m_{j,k,l}^4$.  Performing the
integrals over $\bl^2_{i,\perp}$ with the help of the $\de$ functions,
we find after neglecting irrelevant mass terms
\beqar
\lefteqn{\I_{3\rL} (M_a,M_b;p_j,p_k,p_l) =
\frac{4}{\pi^2} 
(p_j p_k) (p_j p_l) |p_j p_k| |p_j p_l|\int\rd x_1\rd y_1\rd x_2\rd
y_2\rd\phi_1\rd\phi_2}\quad
\nl&&{}
\times\theta(2p_jp_kx_1y_1-M_a^2)
\,\theta(2p_jp_lx_2y_2-M_b^2)
\nl&&{}
\times[2p_jp_kx_1+m_j^2y_1+M_a^2]^{-1}
[-2p_jp_ky_1-m_k^2x_1+M_a^2]^{-1}
\nl&&{}
\times[2p_jp_kx_1+2p_jp_lx_2+m_j^2(y_1+y_2)+M_a^2+M_b^2+2p_kp_lx_1x_2
\nl&&\quad{}
+2p_jp_lx_2y_1+2p_jp_kx_1y_2
+2\bl_{1,\perp}p_lx_2+2\bl_{2,\perp}p_kx_1+2\bl_{1,\perp}\bl_{2,\perp}
]^{-1}
\nl&&{}
\times[-2p_jp_ly_2-m_l^2x_2+M_b^2]^{-1},
\eeqar
where $\bl_{1,\perp}^2=2x_1y_1p_jp_k-M_a^2$ and
$\bl_{2,\perp}^2=2x_2y_2p_jp_l-M_b^2$ and the dependence on $\phi_i$
enters via $\bl_{i,\perp}$.  Terms involving four logarithms result
from those parts of the integration region where all Sudakov variables
are small, $|x_1|,|y_1|,|x_2|,|y_2|\ll 1$, and the integrand behaves
as $1/(x_1y_1x_2y_2)$, \ie where each of the four denominators is
dominated by a different term linear in one of the Sudakov variables.
These parts of the integration region are selected by conditions of
the form $|2p_jp_kx_1|\gg |m_j^2 y_1|$, $|2p_jp_lx_2|\gg |2p_jp_k
x_1|$, etc.  To leading-logarithmic accuracy, these conditions are
implemented via step functions $\theta(|2p_jp_kx_1|-|m_j^2 y_1|)$,
$\theta(|2p_jp_lx_2|-|2p_jp_k x_1|)$, etc., which, in particular,
ensure that none of the Sudakov variables can become zero.  Upper
integration limits are set to one, \ie $|x_1|,|y_1|,|x_2|,|y_2|<1$.
Then irrelevant terms are neglected in the denominators and in the
step functions.  Since, in particular, all terms depending on
$\bl_{i,\perp}$ and thus on $\phi_i$ are negligible, the integration
over these angles can be performed trivially. After transforming
regions with negative Sudakov variables to those with positive Sudakov
variables one finally obtains \refeq{eq:s23sud}.

In this approach it is crucial that the perpendicular components of
the loop momenta $\bl_{i,\perp}$ can be dropped. This is the case for
all ladder diagrams, if the Sudakov parametrizations are constructed
from the external on-shell momenta of the lines to which the gauge
bosons couple as in \refeq{eq:lpdefm}.

The Yang--Mills diagram $\I_{2\rY}$ with two external lines contains
three virtual gauge bosons that can become soft and collinear.  If any
two of them go on-shell, this results in four large logarithms. Thus,
one has to sum over the three contributions with different pairs of
on-shell gauge bosons to obtain the full leading-logarithmic result.
Since there are only two relevant external momenta, the Sudakov
parametrization is unique for each of the three cases.  A slight
complication arises from the fact that the numerator is linear in the
loop momenta.  Nevertheless, one can show that all terms involving
transverse loop momenta can be neglected in the leading-logarithmic
approximation and the leading-logarithmic contributions can be
extracted in a way similar to the ladder diagrams.

For the non-abelian graph $\I_{3\rY}$ with three external lines no
parametrization exists that would allow to neglect all $\bl_{i,\perp}$
terms.  For this diagram we therefore did not apply the Sudakov method
but have checked the result from the sector-decomposition approach by
a numerical integration of the Feynman-parameter integral.

\subsection{Sector-decomposition method}
\label{se:sectordecomp}
The loop integrals were also evaluated in the Feynman-parameter
representation.  In this case, in order to extract the logarithmic
mass singularities we used the sector decomposition
\cite{Hepp:1966eg,Roth:1996pd,Binoth:2000ps}, which permits to
factorize overlapping ultraviolet or mass singularities in
Feynman-parameter integrals. A detailed description of this method is
postponed to a forthcoming publication \cite{nextpaper}.  Here we only
sketch the main steps of the sector-decomposition method applied to a
generic two-loop massive integral with $n+1$ propagators
\beq
\int \frac{\rd^d l_1}{(2\pi)^d}\int\frac{\rd^d l_2}{(2\pi)^d}
\frac{N(\{l_j\},\{p_l\})}{\prod_{i=1}^{n+1}
(k_i^2-m_i^2+\ri \varepsilon)},
\eeq
where the momenta $k_i$ are linear combinations of the external
momenta $p_l$ and the loop momenta $l_j$, and the numerator $N$ is an
arbitrary polynomial in these momenta.

{\bf Step 1:} The integral is written in Feynman-parameter
representation and split into $n+1$ primary sectors as described in
\citere{Binoth:2000ps}.  After eliminating the usual $\de$ function
$\de(1-\sum_{i=1}^{n+1}x_i)$, each primary sector gives rise to a
Feynman-parameter integral over the unit cube in $n$ dimensions of the
form
\beq\label{fprep}
\int_{[0,1]^n} \rd^n \vec{x}\,
\frac{
f(\vec{x})}{\left[
D(\vec{x})
\right]^e}
,\quad
D(\vec{x})=
{s} P_{s}(\vec{x})+{r} P_{r}(\vec{x})+\ldots +{m^2} P_{m}(\vec{x})
+{\la^2} P_{\la}(\vec{x}),
\eeq
where the resulting denominator $D(\vec{x})$ is split into polynomials
according to the hierarchy of scales in the diagram, which is assumed
to be $s\gg r \gg \dots \gg m^2\gg\la^2$ in \refeq{fprep}.  Note that
in order to extract the angular-dependent logarithms $\log{(s/r)}$
with $r=t,u,\dots$, we compute the integrals in the Euclidean region
in various limits of the type $s\gg t=u$, $s=t\gg u$, etc., where we
separate the energy scales in various ways.

{\bf Step 2:}
The polynomials in \refeq{fprep} have various zeros if subsets of
Feynman parameters vanish, \eg
\beq\label{polyone1}
P(\vec{x})=0,
\quad\mbox{at}\quad
x_1=\dots=x_q=0,
\eeq
which give rise to mass singularities.  
In the presence of such a singularity, the polynomial can be written as
\beq\label{polyone2}
P(\vec{x})=\sum_{i=1}^q x_i P_i(\vec{x}).
\eeq
In order to separate the singularity associated to \refeq{polyone1}
from other overlapping singularities, we decompose the corresponding
integration domain $[0,1]^q$ into $q$ subsectors $\Omega_j$ with
$x_j>x_{i\neq j}$, and in each subsector $\Omega_j$ we perform
variable transformations $x_i\to x_j x'_i$ for all $i\neq j$, which
remap $\Omega_j\to[0,1]^q$ and bring the polynomial \refeq{polyone2}
into the form
\beq\label{polytwo}
P(\vec{x})= \left[ P_j(\vec{x})+\sum_{i=1\atop i\neq j}^q x'_i P_i(\vec{x})\right]{x_j}
,
\eeq
where the variable  $x_j$ is factorized.

{\bf Step 3:} Recursive application of step 2 permits to factorize all
zeros at all scales, until the denominator assumes the form%
\footnote{ In general the denominator assumes the form $ D(\vec{x})=
  \hat{D}(\vec{x}) \prod_{i=1}^n {x^{d_i}_i} $, and the overall
  factorized Feynman parameters $x_i$ with $d_i> 0$ can be cancelled
  by corresponding terms in the numerator.}
\beqar\label{eq:den}
\hat{D}(\vec{x})&=&
\left\{\ldots \left[
{s} \hat{P}_s(\vec{x})\prod_{j=1}^n {x^{a_j}_j}
+{r} \hat{P}_r(\vec{x})\right]\prod_{k=1}^n {x_k^{b_k}}
+\ldots 
+{m^2} \hat{P}_m(\vec{x})\right\}\prod_{l=1}^n {x_l^{c_l}}
+{\la^2} \hat{P}_\la(\vec{x})
,\nln
\eeqar
where $a_k,b_k,c_k$ are positive integers.
In \refeq{eq:den} all Feynman parameters that give rise to mass
singularities are factorized and the polynomials $\hat{P}$ are
non-vanishing. This allows for a simple power counting of the
logarithmically divergent integrations.

{\bf Step 4:} All logarithms of ratios of scales can now be extracted
in \NLLad\ approximation by analytical integration, where the
polynomials $\hat{P}$ can be treated as constants
$\hat{P}(\vec{x})\simeq \hat{P}(\vec{0})$.  At present, explicit
results are available \cite{nextpaper} for the special class of
integrals where the various subsets of parameters $\{x_j|a_j>0\}$,
$\{x_k|b_k>0\}$, etc. that are associated to different mass scales are
disjoint.  As an example, for the case of a hierarchy of four
different scales $s\gg r\gg M^2 \gg \la^2$, we have integrals of the
type%
\footnote{ Here we consider all terms with the total power of
  logarithms equal to $N=l+m+n$, but in \NLLad\ approximation we only
  need the contributions with $N-p-q\le 1$.  }
\beqar\label{eq:sd_res}
&&\int_0^1 
\rd^l \vec{x}\,
\int_0^1 
\rd^m \vec{y}\,
\int_0^1 
\rd^n \vec{z}\,
\frac{
{s}^e \left[\prod_{i=1}^l {x_i}
\prod_{j=1}^m {y_j}
\prod_{k=1}^n {z_k}\right]^{e-1}
}{
\left\{
\left[\left(
{s} \prod_{i=1}^l {x_i}
+{r}\right)
\prod_{j=1}^m {y_j}
+{M^2}
\right] 
\prod_{k=1}^n {z_k}
+{\la^2}
\right\}^e
}
\nonumber\\
&&\LA
\sum_{p=0}^n
\sum_{q=0}^{n+m-p}
\frac{1}{p!}
\frac{1}{q!}
\frac{1}{(N-p-q)!}
\log^{p}\left(\frac{{M^2}}{{\la^2}}\right)
\log^{q}\left(\frac{{r}}{{M^2}}\right)
\log^{N-p-q}\left(\frac{{s}}{{r}}\right),
\eeqar
where $N=l+m+n$ and $e\ge 1$.  Such integrals permitted us to
calculate all diagrams listed in \refapp{se:twoloopint} except for the
ladder diagrams with simultaneous photon-mass and external-mass
singularities, since these diagrams lead to integrals where the
various subsets of parameters $\{x_j|a_j>0\}$, $\{x_k|b_k>0\}$, etc.
are not disjoint. Such integrals have not been solved so far,
since it was more convenient to perform the calculation using the
Sudakov method.

\section{Results}
\label{se:results}
In this section we present results for one- and two-loop Feynman
diagrams evaluated in the high-energy limit \refeq{masshierarchy}
using the eikonal approximation (eik) and to next-to-leading
logarithmic angular-dependent (\NLLad) accuracy \refeq{LA}.  All
results are split according to \refeq{eq:subtraction} into
contributions corresponding to $\la=M$ and remaining $\Delta$
contributions, which originate from the gap $\la\ll M$ in the gauge
sector.

In \refse{se:oneloop} we first recall the one-loop results
\cite{Denner:2001jv}, we then present in \refse{se:twoloop} our
results for various subsets of two-loop diagrams and for the complete
two-loop corrections, and in \refse{se:exp} we discuss the
exponentiation of these logarithmic corrections.  Explicit results for
the individual one- and two-loop integrals can be found in
\refapp{app:integrals}.

\subsection{One-loop results}
\label{se:oneloop}

Within the 't~Hooft--Feynman gauge, the one-loop LL and
angular-dependent NLL \cite{Denner:2001jv} originate from diagrams
where a gauge boson $V^a=\gamma,\PZ,\PWpm$ is exchanged between two
different on-shell external legs $j\neq k$,
\beq
\de \M^{i_1\dots i_n}_{1(jk)}=
\sum_a\,
\vcenter{\hbox{
\smalldiagramonetwoL}}.
\eeq
In the eikonal approximation, these yield
\beq
\de \M^{i_1\dots i_n}_{1(jk)}\EIK \frac{\alpha}{4\pi}
\M_{0}^{i_1\dots i'_j\dots i'_k\dots i_n}
 \sum_{a} 
I^{a}_{i_j^\prime i_j}
I^{\bar a}_{i_k^\prime i_k}
\Ione(M_a;p_j,p_k),
\eeq
where the integral $\Ione(M_a;p_j,p_k)$ defined in \refeq{eq:s1} can
be decomposed into
\begin{equation}\label{eq:edef1}
\Ione(M_a;p_j,p_k) \LA  \Eone(M_a;m_j)+\Eone(M_a;m_k) + \Rone(M_a;p_j,p_k), 
\end{equation}
with
\begin{eqnarray}\label{eq:edef2}
\Eone(\lambda;m_j) &=&  \frac{1}{2} \log^2 \frac{\lambda^2}{s}
- \frac{1}{2} \log^2 \frac{\lambda^2}{m_j^2},
\qquad
\Eone(M;m_j) 
=\Eone(M) =  \frac{1}{2} \log^2 \frac{M^2}{s},
\nl
\Rone(M_a;p_j,p_k) &=& 2 \log \frac{M_a^2}{s} \log \frac{s}{|2 p_j
  p_k|}.
\label{eq:r1}
\end{eqnarray}
The functions $\Eone$ depend only on the energy scale $s$, on the
internal masses $M_a$, and on the masses $m_{j,k}$ of the external
lines, and do not give rise to correlations between the two external
legs $j,k$ in the sum \refeq{oneloopres}, whereas the function $\Rone$
contains logarithms of $p_jp_k/s$ depending on the angle between the
momenta $p_j$ and $p_k$, but is independent of the external masses.

The complete one-loop correction is obtained by taking the sum over
all pairs of external legs, and the part originating from the
functions $\Eone$ can be simplified by using the charge-conservation
identity \refeq{eq:gc}. This yields
\beqar\label{oneloopres}
\de \M_1^{i_1\dots i_n} &=& 
\frac{1}{2}\sum_{j,k=1\atop k\neq j}^n\de \M^{i_1\dots i_n}_{1(jk)}
\nl
&=&
-\frac{\alpha}{4\pi}\Biggl\{\sum^n_{j=1} 
\M_0^{i_1\dots i_j^\prime\dots i_n} 
\left[ 
\cew_{i_j^\prime i_j}  \Eone(M) 
+ \de_{i'_j i_j} Q_{i_j}^2 \; \Delta \Eone(\lambda;m_j) \right]
\nl&&{}
- \frac{1}{2}
\sum_{j,k=1\atop k\neq j}^n
\M_0^{i_1\dots i_j^\prime\dots i_k^\prime\dots i_n} 
\biggl[ \sum_{a=\ga,Z,W}
I^a_{i_j^\prime i_j} I^{\bar a}_{i_k^\prime i_k} \Rone(M;p_j,p_k) 
\nl&&\qquad{}
+ \de_{i'_j i_j}Q_{i_j}\de_{i'_k i_k}Q_{i_k} \; \Delta \Rone(\lambda;p_j,p_k)
\biggr]\Biggr\},
\eeqar
where $\cew$ represents the electroweak Casimir operator defined in
\refeq{casimir}, $Q_{i_j}$ is the eigenvalue of the charge operator
$I^\gamma_{i'_ji_j}=-Q_{i_j}\de_{i'_ji_j}$, and
\beqar
\Delta \Eone (\la;m_j)&=& \Eone (\la;m_j)-\Eone (M;m_j)=
\log\left(\frac{m_j^2}{s}\right)
\log\left(\frac{\la^2}{M^2}\right)
-\frac{1}{2}\log^2\left(\frac{m_j^2}{M^2}\right)
,\nl
\Delta \Rone(\lambda;p_j,p_k)&=&\Rone(\lambda;p_j,p_k)- \Rone(M;p_j,p_k)=
2\log\left(\frac{\la^2}{M^2}\right)
\log\left(\frac{s}{|2p_jp_k|}\right).
\eeqar

In order to discuss the two-loop results in the next section, it is
useful to rewrite the one-loop result \refeq{oneloopres} in matrix
form.  To this end we introduce the following notation
\beqar\label{compact}
&&\M\equiv \M^{i_1\dots i_n},\quad
\M I^a(k)\equiv \M^{i_1\dots i'_k\dots i_n} I^a_{i'_ki_k},
\nl
&&\M I^a(k)I^b(k)\equiv 
\M^{i_1\dots i''_k\dots i_n} I^a_{i''_ki'_k}I^b_{i'_ki_k},
\quad\mbox{etc.},
\eeqar
where the generators $I^a(k)$ are matrices acting on the indices
corresponding to the external leg $k$ of the matrix element, with
commutation relations
\beq
\left[ I^a(k), I^b(l) \right] =  \de_{kl} \sum_{c=\gamma,Z,W^\pm}
 I^c(k) I^a_{c b}.
\eeq
Using this notation also for other matrices as $\cew(k)$ or $Q(k)$,
the result \refeq{oneloopres} can be rewritten as
\beqar\label{oponeloopres1}
\de \M_1 &=&\M_0\,  \de_{\EW}=\M_0\left( \de_{\sew}+\de_{\sem}\right),
\eeqar
where the complete electroweak ($\EW$) result is split into a
symmetric electroweak (sew) part
\beqar\label{sewpart}
\de_{\sew}&=& \left.\de_\EW\right|_{\la=M}
\\&=&
\frac{\alpha}{4\pi}\left\{
-\frac{1}{2}\sum_{j=1}^n\cew(j) \log^2\frac{M^2}{s}
+\sum_{j,k=1\atop k\neq j}^n \sum_{a=\gamma,\PZ,\PWpm} 
I^a(j) I^{\bar{a}}(k)
\log\frac{s}{|2p_jp_k|}
\log\frac{M^2}{s}
\right\},\nn
\eeqar
which corresponds to the case $\la=M$ and is manifestly
$\SUtwo\times\Uone$ symmetric, and a remaining subtracted
electromagnetic (sem) part
\beqar\label{sempart}
\de_{\sem} &=&\Delta \de_\EW= \left.\de_\EW -\de_\EW\right|_{\la=M}=
\frac{\alpha}{4\pi}
\left\{
-\frac{1}{2}\sum_{j=1}^n Q^2(j)
\left[
2\log\frac{m_j^2}{s}\log{\frac{\la^2}{M^2}}
-\log^2{\frac{m_j^2}{M^2}}
\right]
\right.\nl && \left.{}
+\sum_{j,k=1\atop k\neq j}^n
Q(j) Q(k)
\log\frac{s}{|2p_jp_k|}
\log\frac{\la^2}{M^2}
\right\},
\eeqar
which originates from the gap $\la\ll M$, \ie from the $\Delta$ terms
in \refeq{oneloopres}, and is $\Uone_\elm$ symmetric.

\subsection{Two-loop results}
\label{se:twoloop}
In the following we present results for the two-loop diagrams
\refeq{diagrams} combined into three subsets where the soft--collinear
gauge bosons couple to two, three, or four of the $n$ on-shell external
lines, respectively.

The two-loop results are decomposed into reducible contributions,
which involve products of the one-loop integrals \refeq{eq:edef1},
plus remaining irreducible parts.  This decomposition is based on the
explicit two-loop results in the $\NLLad$ approximation \refeq{LA} 
given in \refapp{se:twoloopint} and the relations given in
\refapp{se:looprelations}.
 
\subsection*{Terms from two external lines}

We begin by considering the contributions 
\beqar\label{diagtwolegs0}
\de \M_{2(jk)}^{i_1\dots i_n}&=&
\sum_{a,b}\left\{
\diagtwoL{jk}{ab}
+\diagtwoC{jk}{ab}
+\sum_c\left[\diagtwoY{jk}{abc}+\diagtwoY{kj}{abc}
\right]\right\},
\eeqar
corresponding to the diagrams 
\beqar\label{diagtwolegs}
&&
\hspace{-3mm}
\sum_{a,b}
\left\{
\vcenter{\hbox{
\smalldiagramtwoL}}
+
\vcenter{\hbox{
\smalldiagramtwoC}}
{}+\sum_{c}
\left[
\vcenter{\hbox{\smalldiagramtwoY}}
+
\vcenter{\hbox{\smalldiagramtwoYinv}}
\hspace{3mm}\right]\right\}
,\nl
\eeqar
\ie diagrams with $n$ on-shell external legs and soft--collinear gauge
bosons $V^a,V^b,V^c=\gamma,\PZ,\PW$ exchanged between only two of
these external lines.  These yield
\beqar\label{eq:sigma2}
\lefteqn{
\de \M_{2(jk)}^{i_1\dots i_n}
=
\left( \frac{\alpha}{4\pi} \right)^2 
\M_{0}^{i_1\dots i'_j\dots i'_k\dots i_n}
\,\sum_{a,b} \left\{ 
\left( I^{\bar{b}}I^{\bar a} \right)_{i_j^\prime i_j}
\left( I^{{b}} I^{{a}} \right)_{i_k^\prime i_k} 
\I_{2\rL} (M_a,M_b;p_j,p_k)
\right.
}\quad&&
\nl&&\left.{}
+
\left( I^{\bar b}I^{\bar a} \right)_{i_j^\prime i_j} 
\left( I^{a}I^{b} \right)_{i_k^\prime i_k} 
\I_{2\rC} (M_a,M_b;p_j,p_k) 
\right.\nl &&\left.{}
+\sum_c\left[
\left( I^{\bar c}I^{\bar a} \right)_{i_j^\prime i_j}
I^{b}_{i_k^\prime i_k} I^a_{bc} 
\I_{2\rY} (M_a,M_b,M_c;p_j,p_k)
+ 
(j\leftrightarrow k)
\right]\right\}
\nl&\LA&
\left( \frac{\alpha}{4\pi} \right)^2
\M_{0}^{i_1\dots i'_j\dots i'_k\dots i_n}
\left\{ \frac{1}{2} \sum_{a,b} 
\left( I^{\bar b}I^{\bar a} \right)_{i_j^\prime i_j}
\left( I^{b}I^{a} \right)_{i_k^\prime i_k}
\left[\Ione (M;p_j,p_k)\right]^2 
\right. \nl &&{}
+ \sum_{a} \left( I^{\bar a} I^\gamma \right)_{i_j^\prime i_j}  \left( I^{a}
I^\gamma \right)_{i_k^\prime i_k} \Ione (M;p_j,p_k) \Delta \Ione (\lambda;p_j,p_k)
\nl &&{} 
+ \frac{1}{2} \left( I^\gamma I^\gamma \right)_{i_j^\prime i_j}
\left( I^\gamma I^\gamma \right)_{i_k^\prime i_k}
\left[ 
\Delta \Ione (\lambda;p_j,p_k) 
\right]^2 
\nl &&{}\left.
+\frac{1}{2} \sum_{a,c} \sum_{h=1\atop h \neq j,k}^n
I^{\bar a}_{i_h^\prime i_h} \left[ I^{\bar c}_{i_j^\prime i_j}
I^\gamma_{i_k^\prime i_k} I^c_{\gamma a} 
\Delta \I_{2\rC} (M,\lambda;p_j,p_k)
+ 
(j\leftrightarrow k)
\right]\right\},
\eeqar
where we have made use of the identities \refeq{eq:gc},
\refeq{eq:subtraction}, \refeq{eq:commut}, \refeq{eq:ladqed},
\refeq{eq:2Yid}, and \refeq{eq:ladid}.

\subsection*{Terms from three external lines}
Here we consider the contributions
\beqar\label{diagthreelegs0}
\de \M_{2(jkl)}^{i_1\dots i_n}
&=&
\sum_{a,b}\left\{
\sum_{\pi(j,k,l)}\diagthreeL{jkl}{ab}
+\sum_c \diagthreeY{jkl}{abc}
\right\},
\eeqar
where we sum over all six permutations $\pi(j,k,l)$ of external lines
$j,k,l$. These contributions correspond to the diagrams
\beqar\label{diagthreelegs}
&&
\hspace{-2mm}
\sum_{a,b}
\left\{\left[
\vcenter{\hbox{\smalldiagramthreeL}}
+
\vcenter{\hbox{\smalldiagramthreeLinv}}
\;\;\right]
+ 
(j \leftrightarrow k) + ( j \leftrightarrow l)
+\sum_{c}
\vcenter{\hbox{\smalldiagramthreeY}}
\hspace{2mm}\right\}
\nl
\eeqar
with exchange of soft--collinear gauge bosons
$V^a,V^b,V^c=\gamma,\PZ,\PW$ between three external on-shell lines,
which yield
\beqar\label{eq:sigma3}
\de \M_{2(jkl)}^{i_1\dots i_n}
&\EIK&
\left( \frac{\alpha}{4\pi} \right)^2
\M_{0}^{i_1\dots i'_j\dots i'_k\dots i'_l\dots i_n}
\, \nl&&{}\times
\Bigg\{  \sum_{\pi(j,k,l)}\sum_{a,b} 
\left( I^bI^{a} \right)_{i_j^\prime i_j}
I^{\bar a}_{i_k^\prime i_k} I^{\bar b}_{i_l^\prime i_l}
\I_{3\rL} (M_a,M_b;p_j,p_k,p_l) 
\nl&&{}
\quad + \sum_{a,b,c} 
I^{\bar a}_{i_j^\prime i_j} I^a_{bc}
I^{b}_{i_k^\prime i_k} I^{\bar c}_{i_l^\prime i_l}
\I_{3\rY}(M_a,M_b,M_c;p_j,p_k,p_l) \Bigg\} 
\nl&\LA&
\left( \frac{\alpha}{4\pi} \right)^2 
\M_{0}^{i_1\dots i'_j\dots i'_k\dots i'_l\dots i_n}
\,
\nl&&{}
\times\sum_{\pi(j,k,l)}\Bigg\{ 
\frac{1}{2} \sum_{a,b}
\left( I^bI^{a} \right)_{i_j^\prime i_j} 
 I^{\bar a}_{i_k^\prime i_k} I^{\bar b}_{i_l^\prime i_l}
\Ione (M;p_j,p_k) 
\Ione (M;p_j,p_l) 
\nl&&{}
+\sum_{a} \left( I^{a} I^\gamma \right)_{i_j^\prime i_j} 
I^{\bar a}_{i_k^\prime i_k}
I^{\gamma}_{i_l^\prime i_l} \Ione (M;p_j,p_k) \Delta \Ione(\lambda;p_j,p_l)
\nl&&{} 
+ \frac{1}{2}  \left( I^\gamma I^\gamma \right)_{i_j^\prime i_j} 
I^\gamma_{i_k^\prime i_k}I^\gamma_{i_l^\prime i_l}  
\Delta \Ione(\lambda;p_j,p_k) \Delta \Ione (\lambda;p_j,p_l)
\nl&&{}
- 
\frac{1}{2} \sum_{a,c} 
I^{\bar c}_{i_j^\prime i_j} I^c_{\gamma a} 
I^{\bar a}_{i_k^\prime i_k}
I^{\gamma}_{i_l^\prime i_l} \Delta \I_{3\rL} (M,\lambda;p_j,p_k,p_l) 
\Bigg\},
\end{eqnarray}
where we used \refeq{eq:subtraction}, \refeq{eq:commut},
\refeq{eq:angladid}, \refeq{eq:antisymmperm}, \refeq{eq:antisymmzero},
and \refeq{eq:ymid}.

\subsection*{Terms from four external lines}
Finally, we have the contributions 
\beqar\label{diagfourlegs}
\de \M_{2(jklm)}^{i_1\dots i_n}
&=&
\sum_{a,b}\diagfourL{jklm}{ab}=
\sum_{a,b}
\vcenter{\hbox{\smalldiagramfourL}}
\hspace{3mm},
\eeqar
originating from gauge bosons coupling to four external legs, which
reduce according to \refeq{forulegint} to simple products of one-loop
integrals
\beqar\label{eq:sigma4}
\de \M_{2(jklm)}^{i_1\dots i_n}
&\EIK&
\left( \frac{\alpha}{4\pi} \right)^2
\M_{0}^{i_1\dots i'_j\dots i'_k\dots i'_l\dots i'_m\dots i_n}
\, \sum_{a,b} 
I^{a}_{i_j^\prime i_j}
I^{\bar a}_{i_k^\prime i_k}
I^{b}_{i_l^\prime i_l}
I^{\bar b}_{i_m^\prime i_m}
\Ione (M_a;p_j,p_k) 
\Ione (M_b;p_l,p_m).\nln
\end{eqnarray}

\subsection*{Complete two-loop correction}
\newcommand{\sumtwo}[3]{{\mathop{\smash{\phantom{'}{\sum}'}%
\vphantom{\sum}}^#1_{#2,#3}}}
\newcommand{\sumthree}[4]{{\mathop{\smash{\phantom{'}{\sum}'}%
\vphantom{\sum}}\limits^#1_{#2,#3,#4}}}
\newcommand{\sumfour}[5]{{\mathop{\smash{\phantom{'}{\sum}'}%
\vphantom{\sum}}\limits^#1_{#2,#3,#4,#5}}}
We now combine the contributions from the above subsets of diagrams
into the complete virtual two-loop correction to an arbitrary process
involving $n$ on-shell external legs as follows
\beqar\label{combinations}
\delta \M_2^{i_1\dots i_n}&=&
\frac{1}{2}\sumtwo{n}{j}{k}\M^{i_1\dots i_n}_{2(jk)}+
\frac{1}{6}\sumthree{n}{j}{k}{l}\M^{i_1\dots i_n}_{2(jkl)}+
\frac{1}{8}\sumfour{n}{j}{k}{l}{m}\M^{i_1\dots i_n}_{2(jklm)},
\eeqar
where we have to sum over all combinations of external legs with
appropriate symmetry factors. The primes indicate that the sums
include only terms with different external legs, \ie
\beq
\sumtwo{n}{j}{k}:= \sum^n_{j,k=1\atop k\neq j},\qquad\quad
\sumthree{n}{j}{k}{l}:= \sum^n_{j,k,l=1\atop k\neq j;l\neq j,k},\qquad\quad
\sumfour{n}{j}{k}{l}{m}:= \sum^n_{j,k,l,m=1\atop k\neq j;l\neq j, k;
m\neq l,j,k}.
\eeq

We first consider the irreducible contributions to
\refeq{combinations}, \ie the contributions from $\Delta
\I_{2\mathrm{C}}$ in \refeq{eq:sigma2} and $\Delta \I_{\mathrm{3L}}$
in \refeq{eq:sigma3}, which could not be expressed as products of
one-loop integrals. These contributions cancel,
\beqar\label{eq:s2}
\lefteqn{
\delta \M^{i_1\dots i_n}_{2,\mathrm{irr.}}
=
\left( \frac{\alpha}{4\pi} \right)^2
\frac{1}{2} 
\sumthree{n}{j}{k}{l}
\M_0^{i_1\dots i_j^\prime\dots i_k^\prime\dots i_l^\prime\dots i_n}
\sum_{a,c} I^c_{\gamma a} I^{\bar c}_{i_j^\prime i_j} I^\gamma_{i_k^\prime i_k}
I^{\bar a}_{i_l^\prime i_l} 
}\quad&&
\nl &&{} \times 
\left[ \Delta \I_{2\rC} (M,\lambda;p_j,p_k)
- \Delta \I_{3\rL}(M,\lambda;p_j,p_l,p_k) \right]
\LA0,
\eeqar
because of \refeq{eq:3L2Cid}. The complete
two-loop correction is thus given by the reducible contributions to
\refeq{eq:sigma2}, \refeq{eq:sigma3}, and \refeq{eq:sigma4}, \ie
contributions from products of one-loop integrals $\Ione$. These yield
\beqar\label{eq:1lexp}
\lefteqn{
\delta \M^{i_1\dots i_n}_2
\LA
\left( \frac{\alpha}{4\pi} \right)^2
\Bigg\{ \frac{1}{8} \sumfour{n}{j}{k}{l}{m} 
\M_0^{i_1\dots i_j^\prime\dots i_k^\prime\dots i_l^\prime\dots i_m^\prime\dots i_n}
\sum_{a,b}
I^a_{i_j^\prime i_j}  I^{\bar a}_{i_k^\prime i_k} 
I^b_{i_l^\prime i_l}  I^{\bar b}_{i_m^\prime i_m}
}\quad&&
\nl&&{}\qquad\times 
\Ione(M_a;p_j,p_k) \Ione(M_b;p_l,p_m)
\nl&&{}
+ 
\sumthree{n}{j}{k}{l} 
\M_0^{i_1\dots i_j^\prime\dots i_k^\prime\dots i_l^\prime\dots i_n} 
\biggl[
\frac{1}{2} 
\sum_{a,b} 
\left( I^b I^{a} \right)_{i_j^\prime i_j}
I^{\bar a}_{i_k^\prime i_k} I^{\bar b}_{i_l^\prime i_l}
\Ione(M;p_j,p_k) \Ione(M;p_j,p_l)
\nl&&{}+ 
\sum_{a}
\left( I^a I^\gamma \right)_{i_j^\prime i_j}
I^{\bar a}_{i_k^\prime i_k} I^\gamma_{i_l^\prime i_l} 
\Ione(M;p_j,p_k) 
\Delta \Ione(\lambda;p_j,p_l)
\nl &&{}
+ 
\frac{1}{2} \left( I^\gamma I^\gamma \right)_{i_j^\prime i_j} 
I^\gamma_{i_k^\prime i_k} I^\gamma_{i_l^\prime i_l}
\Delta \Ione(\lambda;p_j,p_k)\Delta \Ione(\lambda;p_j,p_l)
\biggr]\nl &&{}
+ 
\sumtwo{n}{j}{k} 
\M_0^{i_1\dots i_j^\prime\dots i_k^\prime\dots i_n}
\biggl[\frac{1}{4} 
\sum_{a,b} \left( I^{\bar b} I^{\bar a} \right)_{i_j^\prime i_j}
 \left( I^{b} I^a \right)_{i_k^\prime i_k} 
\left[\Ione(M;p_j,p_k)\right]^2 
\nl&&{}
+
\frac{1}{2}
\sum_{a}
\left( I^{\bar a} I^\gamma \right)_{i_j^\prime i_j} 
\left( I^{a} I^\gamma \right)_{i_k^\prime i_k}
\Ione(M;p_j,p_k) 
\Delta \Ione(\lambda;p_j,p_k)
\nl &&{}
+
\frac{1}{4} 
\left( I^\gamma I^\gamma \right)_{i_j^\prime i_j} 
\left( I^\gamma I^\gamma \right)_{i_k^\prime i_k}
\left[ \Delta \Ione(\lambda;p_j,p_k)\right]^2
\biggr]\Bigg\}
\nl &\LA& 
\frac{1}{2} \left(\frac{\alpha}{4\pi} \right)^2
\sumtwo{n}{j}{k}
\M_0^{i_1\dots i_j^\prime\dots i_k^\prime\dots i_n} 
\biggl[ \sum_{a}
\left( I^a I^{\bar a} \right)_{i_j^\prime i_j} \Eone(M) 
+\left( I^\gamma I^\gamma \right)_{i'_j i_j} \; \Delta \Eone(\lambda;m_j) \biggr]
\nl &&{} \quad\times 
\biggl[ \sum_b
\left( I^b I^{\bar b} \right)_{i_k^\prime i_k} \Eone(M) 
+ \left( I^\gamma I^\gamma \right)_{i'_k i_k}\; \Delta \Eone(\lambda;m_k) \biggr]
\nl&&{}
+\frac{1}{2} \left(\frac{\alpha}{4\pi} \right)^2
\sum_{j=1}^{n}\M_0^{i_1\dots i_j^\prime\dots i_n}\biggl[ \sum_{a,b}
\left( I^a I^{\bar a} I^b I^{\bar b} \right)_{i_j^\prime i_j} 
[\Eone(M)]^2 
\nl&&\quad{}
+2\sum_{a}
\left( I^a I^{\bar a} I^{\ga} I^{\ga} \right)_{i_j^\prime i_j}
\Eone(M) \Delta \Eone(\lambda;m_j)
+\left( I^{\ga} I^{\ga} I^{\ga} I^{\ga} \right)_{i_j^\prime i_j}
[\Delta \Eone(\lambda;m_j)]^2 
\biggr] 
\nl&&{} 
- \frac{1}{2} \left( \frac{\alpha}{4\pi} \right)^2
\sumthree{n}{j}{k}{l}
\M_0^{i_1\dots i_j^\prime\dots i_k^\prime\dots i'_l\dots i_n}
\biggl[ \sum_b
\left( I^b I^{\bar b} \right)_{i_l^\prime i_l} \Eone(M)
 + \left(I^\gamma I^\gamma\right)_{i'_l i_l} \; 
\Delta \Eone(\lambda;m_l) \biggr]
\nl&&{} \quad\times 
\biggl[ \sum_{a}
I^a_{i_j^\prime i_j} I^{\bar a}_{i_k^\prime i_k} \Rone(M;p_j,p_k) 
+ I^\gamma_{i'_j i_j} I^\gamma_{i'_k i_k} \; \Delta \Rone(\lambda;p_j,p_k)
\biggr]
\nl&&{} 
- \left( \frac{\alpha}{4\pi} \right)^2
\sumtwo{n}{j}{k}
\M_0^{i_1\dots i_j^\prime\dots i_k^\prime\dots i_n}
\biggl[ \sum_{a,b}
(I^a I^b I^{\bar b})_{i_j^\prime i_j} I^{\bar a}_{i_k^\prime
  i_k} 
\, \Eone(M)\,\Rone(M;p_j,p_k) 
\nl&&\quad{}
+ \sum_{a}(I^b I^{\bar b}I^\ga)_{i_j^\prime i_j} I^{\ga}_{i_k^\prime i_k} 
\, \Eone(M)\,\Delta\Rone(\la;p_j,p_k) 
\nl&&\quad{}
+ \sum_{a}(I^a I^\ga I^{\ga})_{i_j^\prime i_j} I^{\bar a}_{i_k^\prime
  i_k} 
\,\Delta \Eone(\lambda;m_j)\,\Rone(M;p_j,p_k) 
\nl&&\quad{}
+(I^\ga I^\ga I^{\ga})_{i_j^\prime i_j} I^{\ga}_{i_k^\prime i_k} 
\,\Delta \Eone(\lambda;m_j)\, \Delta \Rone(\la;p_j,p_k) 
\biggr],
\eeqar
where we have used \refeq{eq:gc}, \refeq{eq:edef1}, \refeq{eq:commut},
\refeq{eq:genid1}, and the fact that $\Rone(M_a;p_j,p_k)$
is symmetric with respect to interchanging $j \leftrightarrow k$.
Note that the photonic coupling matrices $I^\ga I^\ga$ are to the
right of $I^a$ in the term containing $\Delta\Eone(\lambda;m_j)\,
\Rone(M;p_j,p_k)$.

Using $I^\gamma_{i'_ji_j}=-Q_{i_j}\de_{i'_ji_j}$ and rewriting
\refeq{eq:1lexp} in the compact operator form introduced in
\refeq{compact}, results in
\beq\label{matformresult}
\de \M_2 \LA\frac{1}{2}
\M_{0} \left(\de_\sew^2 +2\de_\sew \de_\sem+\de_\sem^2\right), 
\eeq
where $\de_\sew$ and $\de_\sem$ are the operators defined in
\refeq{sewpart} and \refeq{sempart} and correspond to the symmetric
electroweak and the subtracted electromagnetic parts of the one-loop
corrections.  It is important to note that these two operators have a
non-vanishing commutator
\beq\label{commutator}
[\de_\sew, \de_\sem]=\ord\left[
\log\frac{|2p_{k}p_{l}|}{s}
\log^{3-N}\frac{s}{M^2}
\prod_{i=1}^N\log\frac{M^2}{m_{\mathrm{light},i}^2}
\right],\quad N=1,2,
\eeq
where $m_{\mathrm{light}}$ is either a light-fermion or a photon mass.
This means that the ordering of the terms $\de_\sew\de_\sem$ in
\refeq{matformresult}, which is determined by the contribution
involving $I^aI^\ga I^\ga$ in \refeq{eq:1lexp}, is relevant at the
level of angular-dependent $\NLL$.  As we point out in the next
section, the determination of this ordering constitutes an important
aspect of our result and permits to discriminate between different
exponentiation prescriptions for the electroweak corrections.
 
\subsection{Exponentiation}
\label{se:exp}
If we combine the two-loop correction \refeq{matformresult} with the
one-loop correction \refeq{oponeloopres1} and the Born amplitude, to
two-loop $\NLLad$ accuracy we find the exponentiated form
\beq\label{result}
\M_2=\M_0+\de \M_1+\de \M_2 
\LA
\M_{0}
\exp\left(\de_\sew\right) 
\exp\left(\de_\sem\right).
\eeq
In particular, the form of the two-loop correction operator
$\left(\de_\sew^2 +2\de_\sew \de_\sem+\de_\sem^2\right)/2$ implies
that the symmetric electroweak part $\de_\sew$ and the subtracted
electromagnetic part $\de_\sem$ exponentiate separately, and that the
latter exponential is external.  This means that the charge operators
in $\exp(\de_\sem)$ can be identified with the charge eigenvalues of
the external particles in the process.

At the level of the LL, this result confirms the exponentiation of the
EWLC obtained with the IREE \cite{Fadin:2000bq} and already checked
for arbitrary processes by a two-loop calculation in the Coulomb gauge
\cite{Beenakker:2000kb}.  We found also agreement with the results of
\citeres{Melles:2000ed,Hori:2000tm} for the special case of the
fermionic form factor corresponding to the decay $g\to f\bar f$ of an
$\SUtwo\times\Uone$ singlet $g$ into massless fermions.  We have
explicitly verified all results of \citeres{Melles:2000ed,Hori:2000tm}
by evaluating the subset of diagrams $\symbdiag_{2{\mathrm L}}$,
$\symbdiag_{2{\mathrm C}}$, and $\symbdiag_{2{\mathrm Y}}$ in
\refeq{diagrams} for the special case of massless external particles.
At the level of angular-dependent NLL, our result is in agreement with
the exponentiation prescriptions adopted in
\citeres{Kuhn:2000nn,Kuhn:2001hz} for massless fermionic processes,
and extended in \citere{Melles:2001dh} to arbitrary processes.
 
This agreement indicates that, at least up to the level of
angular-dependent $\NLL$, a symmetric $\SUtwo\times\Uone$ gauge theory
matched with QED at the electroweak scale provides a correct physical
picture for the resummation of EWLC in the high-energy limit.  This
picture has been formulated within the theoretical framework of the
IREE \cite{Fadin:2000bq}, which describes the all-order
leading-logarithmic dependence of matrix elements on the
transverse-momentum cut-off $\mu_\perp$.  This infrared scale
$\mu_\perp$ is the crucial ingredient in order to avoid the
difficulties related to the breaking of the $\SUtwo\times\Uone$
symmetry that originate from the large mass gap $\la \ll M$ in the
gauge-boson sector.  In fact, the scale $\mu_\perp$ permits to
separate two regimes of the electroweak theory both with exact gauge
symmetry.  The regime $\sqrt{s}>\mu_\perp>M$, which is insensitive to
the gauge-boson masses and has $\SUtwo\times\Uone$ symmetry, and the
regime $M>\mu_\perp>\la$, where the weak gauge bosons are ``frozen
out'' and only $\Uone_\elm$ symmetry is left.

To our knowledge, the IREE has been formulated only at the level of
LL, and the application of the physical picture described above to the
level of NLL relies on a weaker theoretical basis.  At this level, the
following two arguments can be used for the exponentiation of the
next-to-leading logarithmic corrections.  On the one hand, if $\la=M$
then the $\SUtwo\times\Uone$ symmetry is restored in the gauge sector
and one expects the exponentiation
\beqar\label{matching1}
&&
\la=M \;\Rightarrow\; 
\M_2=
 \M_{0}\exp\left[\de_\sew\right],
\eeqar
as in a symmetric $\SUtwo\times\Uone$ theory. This permits to predict
the two-loop term proportional to $\de_\sew^2$ in \refeq{result} and
implies that $\de_\sem =0$ at $\la=M$.  On the other hand, the
logarithms of the photon mass and light-fermion masses originate only
from photons coupling to external legs and are expected to
exponentiate as in QED. However, the QED results can be generalized to
the electroweak corrections only if the contributions from virtual
photons can be separated from those of the weak gauge bosons in a
gauge-invariant way.  This is the case only if $s=M^2$, where the
logarithms of $s/M^2$ originating from virtual weak bosons vanish.
Here one expects
\beqar\label{matching2}
&&s=M^2 \;\Rightarrow\; 
\M_2= \M_{0}\exp\left[\de_\sem\right],
\eeqar
where $\de_\sem$ corresponds to the QED corrections.  This, together
with \refeq{matching1}, permits to determine $\de_\sem$ and the
two-loop term proportional to $\de_\sem^2$ in \refeq{result}.
However, we note that the above two conditions, \refeq{matching1} and
\refeq{matching2}, are not sufficient to determine the two-loop
interference terms $\de_\sew \de_\sem$ between symmetric electroweak
and subtracted electromagnetic contributions, which vanish in both
cases $\la=M$ and $s=M^2$.  These two-loop interference terms are an
important result of our electroweak calculation for $s\gg
M^2\gg\la^2$. In particular, they are crucial in order to predict the
ordering of the two exponentials in \refeq{result}, which starts to be
non-trivial at the level of angular-dependent NLL as indicated by the
commutator \refeq{commutator}.

\section{Conclusions}

We have studied the two-loop asymptotic behaviour of virtual
electroweak corrections to arbitrary processes involving light or
heavy chiral fermions, transverse or longitudinal gauge bosons, or
Higgs bosons. We have calculated the two-loop leading and
angular-dependent next-to-leading logarithmic contributions in a
process-independent way in the region where all kinematic invariants
are much larger than the electroweak scale.  The relevant Feynman
diagrams involving exchanges of soft and collinear virtual gauge
bosons $\gamma$, $\PZ$, and $\PW^\pm$ between on-shell external legs
have been evaluated in the eikonal approximation in the
't~Hooft--Feynman gauge.

Analytical expressions for the relevant two-loop integrals, which
involve up two six different scales, have been calculated via two
independent methods.
On the one hand, we have evaluated the integrals in the
Feynman-parameter representation using sector decomposition to isolate
the mass singularities in the integrand and performing the integration
in logarithmic approximation.  This method was applied to all diagrams
except for those ladder diagrams with simultaneous photon-mass and
external-mass singularities.
On the other hand, we have employed the well-known Sudakov method,
which is very efficient for calculating diagrams with only one large
energy scale but turns out to be more complicated
for diagrams with more
large energy scales. In particular, in the Sudakov approximation we
did not succeed in calculating the diagram where three soft gauge
bosons interacting via a Yang--Mills vertex couple to three different
external legs.
In all diagrams where both methods could be applied we found agreement
at the angular-dependent
next-to-leading logarithmic level.

In order to isolate the effects originating from the large mass-gap
between the photon mass $\la$ and the weak-boson masses
$\MW\simeq\MZ\simeq M$, which breaks the symmetry in the gauge-boson
sector, the loop contributions depending on the photon mass have been
split into a part corresponding to $\la=M$ and a remaining subtracted
part.
Combining the results from all diagrams we found that the sum of the
two-loop leading and angular-dependent next-to-leading logarithmic
corrections can be written as the second-order term of a product of
two exponential functions.  The first exponential contains the part of
the corrections corresponding to $\la=M$, \ie the $\SUtwo\times\Uone$
symmetric part. The second, outer exponential contains the
contributions that originate from the mass gap $\la\ll M$ and
corresponds to the QED corrections subtracted in such a way that they
vanish at $\la= M$.

This result agrees with resummation prescriptions that have been
proposed in the literature. These prescriptions are based on the
assumption that, in the high-energy limit, the electroweak theory can
be described by a symmetric $\SUtwo\times\Uone$ theory matched with
QED at the electroweak scale, and that no additional effects from
spontaneous symmetry breaking appear.  Our result, which has
been derived within the spontaneously broken phase of the electroweak
theory and in the physical basis, demonstrates that this assumption is
correct at the next-to-leading angular-dependent logarithmic level.

Our derivations depend only on a few general features of the
Electroweak Standard Model, \ie on the underlying gauge symmetry and
the fact that all particle masses are of the order of the electroweak
scale or lighter.
Therefore our result is also valid for those extensions of the Electroweak 
Standard Model that contain only novel particles with masses of the
order of the weak scale and no additional gauge bosons. 
%
Such models include, for instance, the Electroweak Standard Model with
two Higgs doublets or softly broken supersymmetric extensions such as
the Minimal Supersymmetric Standard Model in the case where the masses
of the Higgs bosons and the superpartner particles are of the order of
the electroweak scale.

\section*{Acknowledgements}
This work was supported in part by the Swiss Bundesamt f\"ur Bildung
und Wissenschaft and by the European Union under contract
HPRN-CT-2000-00149. We thank Stefan Dittmaier for carefully reading
the manuscript.

\begin{appendix}
\section{Gauge-group generators}
\label{se:couplings}
All our derivations are performed in terms of the physical
(mass-eigenstate) gauge bosons $\gamma,\PZ,\PWpm$.  The corresponding
gauge couplings result from combinations of the generators $T^a$ and
$Y$ of the electroweak ${\SUtwo}\times{\Uone}$
gauge group, and read
\beq
\label{eq:gendef}
I^\gamma=-Q,\qquad I^Z=\frac{T^3-\sw^2 Q}{\sw\cw},\qquad
I^{W^\pm}=\frac{1}{\sw}T^\pm=\frac{1}{\sw}\frac{T^1\pm\ri T^2}{\sqrt{2}},
\eeq
where $Q=T^3+Y/2$ represents the electric charge, and we use the
shorthands $\cw=\cos{\thw}$ and $\sw=\sin{\thw}$ for the weak mixing
angle, which is fixed by $\cw^2=1-\sw^2=\MW^2/\MZ^2$ in the on-shell
renormalization scheme.

For the matrix components of the generators \refeq{eq:gendef} we use
the notation $I^a_{i'i}$, where $a=\gamma,Z,W^\pm$ denotes the gauge
fields, whereas the indices $i'$ and $i$ correspond to two physical
(mass-eigenstate) components $\varphi_{i'}$ and $\varphi_{i}$ of a
multiplet.  The explicit matrix representations corresponding to the
scalar doublet, right- or left-handed fermions and gauge bosons, as
well as more details concerning our conventions, can be found in
App.~B of \citere{Pozzorini:rs}.

The matrix component $I^a_{i'i}$ determines the gauge coupling for the
vertex with the particles $V^a$ and $\varphi_i$ incoming and the
particle $\varphi_{i'}$ outgoing. The fields and the matrix components
are in general complex and satisfy the relations \cite{Pozzorini:rs}
\beqar
\left(I^{a}_{j i}\right)^*=-I^{\bar{a}}_{\bar{j}\bar{i}},\qquad
I^{a}_{ij}
=-I^{a}_{\bar{j}\bar{i}},
\eeqar
where the particles $\bar{a},\bar{i},\dots$ correspond to the charge
conjugated of $a,i,\dots$.

In our derivations we make extensive use of the commutation relations 
\begin{equation}\label{structurefun}
\left[ I^a, I^b \right] =  \sum_{c=\gamma,Z,W^\pm}
I^c I^a_{c b}
\label{eq:commut}
\end{equation}
and of the well-known commutation relations
\beqar
 \sum_{a=\gamma,Z,W^\pm} \left[ I^a I^{\bar a}, I^b \right] &=& 0
\qquad \mbox{with}\quad b=\gamma,Z,W^\pm,
\label{eq:genid1}
\eeqar
for the electroweak Casimir operator
\beq\label{casimir}
\cew :=\sum_{a=\gamma,Z,W^\pm} I^a I^{\bar a}
=\frac{1}{\cw^2}\left(\frac{Y}{2}\right)^2+\frac{1}{\sw^2} 
T(T+1),
\eeq
where $T$ is the total isospin, and $T(T+1)$ is the Casimir operator
of the SU(2) group.  The electroweak Casimir operator is a diagonal
matrix apart from the neutral gauge-boson sector, where mixing gives
rise to the non-diagonal components $\cew_{\gamma Z}=\cew_{Z\gamma}=
-2\cw/\sw$ 
(see App.~B of \citere{Pozzorini:rs}).

\section{1- and 2-loop integrals in logarithmic approximation}
\label{app:integrals}

In this section we present detailed results for the one- and two-loop
integrals involving soft--collinear gauge bosons.  For each diagram we
first specify the corresponding Feynman integral in eikonal
approximation (eik.). Then we also give the corresponding integral in
the Sudakov approximation (Sud.).
Finally, we present explicit results in next-to-leading logarithmic
angular-dependent (\NLLad)
approximation \refeq{LA}, which have been obtained in the high-energy
limit \refeq{masshierarchy} and for all cases specified in \refeq{smcase2}
and \refeq{smcase3}.  These results were derived using the Sudakov
approximation and the sector-decomposition method described in
\refse{se:sudakov} and \refse{se:sectordecomp}, respectively.

The external momenta are assumed to be on-shell, $p_k^2=m_k^2$, with
masses at or below the electroweak scale, \ie $M\gsim m_k\gg\la$. 
Masses that do not regularize mass singularities are neglected.
Consequently, the masses of the external particles and of the internal
particles that are not soft--collinear gauge bosons are only
relevant for photon exchange diagrams, where the masses before and
after photon emission are equal.  Therefore, we can set the internal
and external masses of the particle lines equal in the following.

\subsection{One-loop integrals}\label{se:oneloopint}

For the one-loop diagram 
\beqar\label{2legsoneladder}
\vcenter{\hbox{
\eikdiagramonetwoL
}} 
&\EIK&
\frac{\alpha}{4\pi}\Ione(M_a;p_j,p_k)
\,
\M_0^{i_1\dots i'_j\dots i'_k\dots i_n}
I^a_{i'_j i_j}
I^{\bar{a}}_{i'_k i_k}
\eeqar
we have the integral
\beqar\label{eq:s1}
\Ione(M_a;p_j,p_k) &:=& - {\ri (4 \pi)^2 \int} \frac{{\rd^4}{l_1}}{(2 \pi)^4} 
\frac{ 4 p_j p_k}{[l_1^2-M_a^2]
[(p_j-l_1)^2-m_j^2][(p_k+l_1)-m_k^2]}.
\eeqar
Here and in the following we suppress the infinitesimal imaginary
parts $\ri\veps$ of the causal propagators for brevity.
In the Sudakov approximation 
\beqar\label{eq:s1sud}
\lefteqn{
\Ione(M_a;p_j,p_k)\SUD
}\quad&&
\\ &=&
2 \int^1_0 \frac{\rd x_1}{x_1} \int^1_0 \frac{\rd y_1}{y_1} \,
\theta\! \left( x_1y_1- \frac{M_a^2}{|2p_jp_k|} \right)
\theta\! \left( y_1 - \frac{m_j^2}{|2p_jp_k|} x_1 \right)
\theta\! \left( x_1 - \frac{m_k^2}{|2p_jp_k|} y_1 \right).\nn 
\eeqar
The results in \NLLad\ approximation corresponding to the cases
$M_a=M,\la$, are given in \refeq{eq:edef1} and \refeq{eq:edef2}.

\subsection{Two-loop integrals}\label{se:twoloopint}

\subsubsection*{2-leg ladder diagram $\I_{2\rL}$}

We begin with the planar ladder diagram
\beqar\label{2legsladder}
\diagtwoL{jk}{ab}:=
\vcenter{\hbox{
\eikdiagramtwoL
}} 
&\EIK&
\begin{array}{l}
\\ \displaystyle{
\left(\frac{\alpha}{4\pi}\right)^2
\I_{2\rL}(M_a,M_b;p_j,p_k)
}\\ \displaystyle{\hspace{0mm}{}\times
\M_0^{i_1\dots i'_j\dots i'_k\dots i_n}
(I^b I^a)_{i'_j i_j}
(I^{\bar{b}} I^{\bar{a}})_{i'_k i_k}
}\end{array}
\eeqar
with 
\begin{eqnarray}\label{eq:s21}
\lefteqn{\I_{2\rL}(M_a,M_b;p_j,p_k) := 
- (4\pi)^4 \! \int \!\! \frac{\rd^4l_1}{(2 \pi)^4} \!\! \int \!\!
\frac{\rd^4l_2}{(2 \pi)^4}
 \frac{1}{[l_1^2-M_a^2][(p_j-l_1)^2-m_j^2]}}\quad&&
\nl&&{}\times 
\frac{16(p_j p_k)^2}{[(p_k+l_1)^2-m_k^2]
[l_2^2-M_b^2][(p_j-l_1-l_2)^2-m_j^2][(p_k+l_1+l_2)^2-m_k^2]}.
\end{eqnarray}
In the Sudakov approximation
\begin{eqnarray}\label{eq:s21sud}
\lefteqn{\I_{2\rL}(M_a,M_b;p_j,p_k) \SUD
}\quad&&
\nl
&=&
4 \! \int^1_0 \frac{\rd x_1}{x_1} \! \int^1_0 \frac{\rd y_1}{y_1} 
\! \int^1_0 \frac{\rd x_2}{x_2}
\! \int^1_0 \frac{\rd y_2}{y_2}\,  
\theta\!\left( x_1y_1- \frac{M_a^2}{|2p_jp_k|} \right)
\theta\! \left( x_2y_2- \frac{M_b^2}{|2p_jp_k|} \right) 
\nl &&{}\times
\theta\! \left( x_2-x_1 \right) \theta \left( y_2-y_1 \right) 
\theta\! \left( x_1-\frac{m_j^2}{|2p_jp_k|}y_1 \! \right)
\theta\! \left( y_1-\frac{m_k^2}{|2p_jp_k|}x_1 \! \right)
\nl && {}\times 
 \theta \! \left( x_2-\frac{m_j^2}{|2p_jp_k|}y_2 \! \right) 
\theta \! \left( y_2-\frac{m_k^2}{|2p_jp_k|}x_2 \! \right).
\end{eqnarray}
In \NLLad\ approximation, we find the following expressions for the
cases \refeq{smcase2}:
\begin{eqnarray}
\lefteqn{
\I_{2\rL}(\lambda,\lambda;p_j,p_k) 
\LA 
\frac{1}{2} \log^2 \frac{\lambda^2}{|2p_jp_k|}
\log^2 \frac{m_j^2m_k^2}{(2p_jp_k)^2}
-\frac{1}{6} \log \frac{\lambda^2}{|2p_jp_k|} 
}\quad&&\nl 
&&\quad\times
\left[ 2 \log^3 \frac{m_j^2m_k^2}{(2p_jp_k)^2}
+ 3 \log \frac{m_j^2m_k^2}{(2p_jp_k)^2}
\left( \log^2 \frac{m_j^2}{|2p_jp_k|}+\log^2 \frac{m_k^2}{|2p_jp_k|} \right) \right]
\nl&&{}
+ \frac{5}{12} \left[\log^4 \frac{m_j^2}{|2p_jp_k|}+\log^4 \frac{m_k^2}{|2p_jp_k|}\right]
+\log^2 \frac{m_j^2}{|2p_jp_k|} \log^2 \frac{m_k^2}{|2p_jp_k|} 
\nl &&{}
+\frac{5}{6} \log \frac{m_j^2}{|2p_jp_k|} \log \frac{m_k^2}{|2p_jp_k|} \left( \log^2 \frac{m_j^2}{|2p_jp_k|}
+ \log^2 \frac{m_k^2}{|2p_jp_k|}\right), 
\label{eq:s21ll} \\
\lefteqn{
\I_{2\rL}(\lambda,M;p_j,p_k) 
\LA 
- \frac{1}{6} \log^4\frac{M^2}{|2p_jp_k|}
-\frac{2}{3} \log^3 \frac{M^2}{|2p_jp_k|} \log \frac{m_j^2m_k^2}{(2p_jp_k)^2} 
}\quad&&
\nl&&{}  
- \frac{1}{2} \log^2 \frac{M^2}{|2p_jp_k|} 
\left[ \log^2 \frac{m_j^2}{|2p_jp_k|}+
\log^2 \frac{m_k^2}{|2p_jp_k|} -2 \log \frac{\lambda^2}{|2p_jp_k|} \log
\frac{m_j^2m_k^2}{(2p_jp_k)^2} \right],  
\label{eq:s21ml}\\ 
\lefteqn{\I_{2\rL}(M,\lambda;p_j,p_k) 
\LA
\I_{2\rL}(M,M;p_j,p_k) \LA 
\frac{1}{6}\log^4 \frac{M^2}{|2p_jp_k|}.
}\quad 
\label{eq:s21mm}
\end{eqnarray}

\subsubsection*{2-leg crossed ladder diagram $\I_{2\rC}$}

For the 2-leg crossed (non-planar) ladder diagram
\beqar\label{2legscrossedladder}
\diagtwoC{jk}{ab}:=
\vcenter{\hbox{
\eikdiagramtwoC
}} 
&\EIK&
\begin{array}{l}
\\ \displaystyle{
\left(\frac{\alpha}{4\pi}\right)^2
\I_{2\rC}(M_a,M_b;p_j,p_k)
}\\ \displaystyle{\hspace{0mm}{}\times
\M_0^{i_1\dots i'_j\dots i'_k\dots i_n}
(I^b I^a)_{i'_j i_j}
(I^{\bar{a}} I^{\bar{b}})_{i'_k i_k}
}\end{array}
\eeqar
we have
\begin{eqnarray}\label{eq:s22}
\lefteqn{
\I_{2\rC}(M_a,M_b;p_j,p_k) := - (4\pi)^4 \int \frac{\rd^4l_1}{(2 \pi)^4} \int \frac{\rd^4l_2}{(2 \pi)^4}
\frac{1}{[l_1^2-M_a^2][(p_j-l_1)^2-m_j^2]}
}\quad&&
\nl&&{}
\times \frac{16[p_j(p_k-l_2)][p_k(p_j-l_1)]}{
[l_2^2-M_b^2][(p_k-l_2)^2-m_k^2][(p_j-l_1+l_2)^2-m_j^2]
[(p_k+l_1-l_2)^2-m_k^2]}.
\nln
\end{eqnarray}
As discussed in \refse{se:eikonal}, in order to avoid spurious leading
logarithms originating from the region $l_1\approx p_j$ and
$l_2\approx p_k$ when the integral is evaluated in the
Feynman-parameter representation, loop-momentum-dependent eikonal
factors \refeq{eikvertexb} have to be used for the inner vertices of
this topology.

In the Sudakov approximation, where the loop-momentum dependence of
the eikonal factors can be neglected, we have
\begin{eqnarray}\label{eq:s22sud}
\lefteqn{
\I_{2\rC}(M_a,M_b;p_j,p_k) \SUD
}\quad&&
\nl&=& 
4 \int^1_0 \frac{\rd x_1}{x_1} \int^1_0 \frac{\rd y_1}{y_1} \int^1_0 \frac{\rd x_2}{x_2}
\int^1_0 \frac{\rd y_2}{y_2}\, 
\theta\! \left( x_1y_1- \frac{M_a^2}{|2p_jp_k|} \right)
\theta\! \left( x_2y_2- \frac{M_b^2}{|2p_jp_k|} \right)
\nl &&
{}\times\theta\! \left( y_1-\frac{m_j^2}{|2p_jp_k|}x_1 \right)
\theta\! \left( x_2-\frac{m_k^2}{|2p_jp_k|}y_2 \right) 
\theta\! \left( y_2-y_1 \right) \theta(x_1-x_2). 
\end{eqnarray}
In \NLLad\ approximation, we derive the following expressions for the
cases \refeq{smcase2}:
\begin{eqnarray}\label{eq:s22ll} 
\lefteqn{
\I_{2\rC}(\lambda,\lambda;p_j,p_k) \LA  \frac{1}{3} \log
\frac{\lambda^2}{|2p_jp_k|} \log^3 \frac{m_j^2m_k^2}{(2p_jp_k)^2}
}\quad&&
\nl&&{}
-\frac{7}{24} \left( \log^4\frac{m_j^2}{|2p_jp_k|}
+\log^4 \frac{m_j^2}{|2p_jp_k|} \right) 
-\frac{3}{4} \log^2 \frac{m_j^2}{|2p_jp_k|} \log^2  \frac{m_k^2}{|2p_jp_k|}
\nl&&{}
 - \frac{5}{6}\log \frac{m_j^2}{|2p_jp_k|} \log\frac{m_k^2}{|2p_jp_k|} 
\left[ \log^2 \frac{m_j^2}{|2p_jp_k|}+ \log^2
\frac{m_k^2}{|2p_jp_k|} \right]
,
\\
\lefteqn{
\I_{2\rC}(\lambda,M;p_j,p_k) =
\I_{2\rC}(M,\lambda;p_k,p_j) \LA 
\frac{2}{3} \log^3\frac{M^2}{|2p_jp_k|} \log \frac{m_j^2}{|2p_jp_k|}, \label{eq:s22ml} 
}\quad&&
\\
\lefteqn{
\I_{2\rC}(M,M;p_j,p_k) \LA \frac{1}{3} \log^4 \frac{M^2}{|2p_jp_k|}. \label{eq:s22mm}
}\quad&&
\end{eqnarray}

\subsubsection*{2-leg Yang--Mills diagram $\I_{2\rY}$}

For the 2-leg Yang--Mills diagram 
\beqar\label{2legsym}
\diagtwoY{jk}{abc}:=
\vcenter{\hbox{
\eikdiagramtwoY
}} 
&\EIK&
\begin{array}{l}
\\ \displaystyle{
\left(\frac{\alpha}{4\pi}\right)^2
\I_{2\rY}(M_a,M_b,M_c;p_j,p_k)
}\\ \displaystyle{\hspace{0mm}{}\times
\M_0^{i_1\dots i'_j\dots i'_k\dots i_n}
(I^{\bar c} I^{\bar a})_{i'_j i_j}
I^{b}_{i'_k i_k}
I^a_{bc}
}\end{array}
\eeqar
we have 
\begin{eqnarray}\label{eq:s31}
\lefteqn{
\I_{2\rY} (M_a,M_b,M_c;p_j,p_k)
:=  (4\pi)^4 \int \frac{\rd^4l_1}{(2 \pi)^4} \int \frac{\rd^4l_2}{(2 \pi)^4}
\frac{1}{ [l_1^2-M_a^2][(p_j-l_1)^2-m_j^2]}
}\quad&&
\nl&&{} \times
\frac{8p_jp_k [(l_2+2 l_1) p_j]}{
[l_2^2-M_b^2][(l_1+l_2)^2-M_c^2] [(p_j+l_2)^2-m_j^2] [(p_k-l_2)^2-m_k^2]}.
\end{eqnarray}
For the cases \refeq{smcase3} we find the Sudakov approximation
\begin{eqnarray}\label{eq:s31sud}
\lefteqn{
\I_{2\rY} (M_a,M_b,M_c;p_j,p_k)\SUD
}\quad&&
\nl
&=&
2 \int^1_0 \frac{\rd x_1}{x_1} \int^1_0 \frac{\rd y_1}{y_1} \int^1_0 \frac{\rd x_2}{x_2}
\int^1_0 \frac{\rd y_2}{y_2}\, \theta\! \left( x_1y_2- x_2y_1 \right)
\theta\! \left( y_1- \frac{m_j^2}{|2p_jp_k|}x_1 \right)
\nl&&{} \times 
\theta\! \left( y_2-\frac{m_j^2}{|2p_jp_k|} x_2 \right) 
\theta\! \left( x_2-\frac{m_k^2}{|2p_jp_k|} y_2 \right) 
\nl&&{}\times
\Biggl\{ \theta\!\left( x_1 y_1 - \frac{M_c^2}{|2p_jp_k|} \right)
\theta\! \left( x_2 y_2 -\frac{M_b^2}{|2p_jp_k|} \right) 
\theta\! \left( y_1-y_2 \right)
+ \theta\! \left( x_1 y_1 - \frac{M_a^2}{|2p_jp_k|} \right)
\nl&&{}\times
\theta\! \left( x_2 y_2 -\frac{M_c^2}{|2p_jp_k|} \right) 
\theta\!\left(x_2-x_1 \right) 
\theta\! \left(y_2-y_1 \right) 
- \theta\! \left( x_1 y_1 - \frac{M_a^2}{|2p_jp_k|} \right)
\nl&&{}\times
\theta\! \left( x_2 y_2 - \frac{M_b^2}{|2p_jp_k|} \right) 
\theta\! \left( x_1 y_2- \left| \frac{M_c^2}{|2p_jp_k|}-\frac{M_a^2}{|2p_jp_k|}-\frac{M_b^2}{|2p_jp_k|}\right| \right)
\Biggr\},
\end{eqnarray}
and in \NLLad\ approximation we obtain
\beqar
\I_{2\rY} (\la,M,M;p_j,p_k)&\LA& 
\I_{2\rY} (M,\la,M;p_k,p_j)\LA 
\nl&\LA&{}
-\frac{1}{3} \log^3\frac{M^2}{|2p_jp_k|} \log \frac{m_j^2}{|2p_jp_k|},
\label{eq:s311res}
\\
\I_{2\rY} (M,M,\la;p_j,p_k)&\LA& 
\I_{2\rY} (M,M,M;p_j,p_k)\LA 
-\frac{1}{6} \log^4 \frac{M^2}{|2p_jp_k|}.
\label{eq:s312res}
\eeqar

\subsubsection*{3-leg ladder diagram $\I_{3\rL}$}

For the 3-leg ladder diagram
\beqar\label{3legsladder}
\diagthreeL{jkl}{ab}:=
\vcenter{\hbox{
\eikdiagramthreeL
}} 
\quad&\EIK&
\begin{array}{l}
\\ \displaystyle{
\left(\frac{\alpha}{4\pi}\right)^2
\I_{3\rL}(M_a,M_b;p_j,p_k,p_l)
}\\ \displaystyle{\hspace{0mm}{}\times
\M_0^{i_1\dots i'_j\dots i'_k\dots i'_l\dots i_n}
I^{a}_{i'_k i_k}
(I^b I^{\bar{a}})_{i'_j i_j}
I^{\bar{b}}_{i'_l i_l}
}\end{array}
\eeqar
we have
\begin{eqnarray}\label{eq:s23}
\lefteqn{
\I_{3\rL} (M_a,M_b;p_j,p_k,p_l) := - (4\pi)^4
\int \frac{\rd^4l_1}{(2 \pi)^4} \int \frac{\rd^4l_2}{(2 \pi)^4}
\frac{1}{[l_1^2-M_a^2][(p_j+l_1)^2-m_j^2]}
}\quad&&
\nl&&{}\times 
\frac{16 (p_j p_k)[(p_j+l_1)p_l]}{
[(p_k-l_1)^2-m_k^2][l_2^2-M_b^2][(p_j+l_1+l_2)^2-m_j^2]
[(p_l-l_2)^2-m_l^2]}.
\end{eqnarray}
As discussed in \refse{se:eikonal}, in order to avoid spurious leading
logarithms originating from the region $l_1\approx -p_j$ and
$l_2\approx 0$ when the integral is evaluated in the Feynman-parameter
representation, a loop-momentum-dependent eikonal factor
\refeq{eikvertexb} has to be used for the emission of the gauge boson
$V^b$ along the line $j$ in this topology.

In the Sudakov approximation, where the loop-momentum dependence of
the eikonal factor can be neglected, we have
\begin{eqnarray}\label{eq:s23sud}
\lefteqn{
\I_{3\rL} (M_a,M_b;p_j,p_k,p_l)\SUD
}\quad&&
\nl &=&
4 \int^1_0 \frac{\rd x_1}{x_1} \int^1_0 \frac{\rd y_1}{y_1} \int^1_0 \frac{\rd x_2}{x_2}
\int^1_0 \frac{\rd y_2}{y_2} \,
\theta\! \left( x_1y_1- \frac{M_a^2}{|2p_jp_k|} \right)
\theta\! \left( x_2y_2- \frac{M_b^2}{|2p_jp_l|} \right)  
\nl&&{}\times
\theta\! \left( x_2-  \frac{|p_jp_k|}{|p_jp_l|}x_1 \right) 
\theta\! \left( x_1-\frac{m_j^2}{|2p_jp_k|}y_1 \right)
\theta\! \left( x_2-\frac{m_j^2}{|2p_jp_l|}y_2 \right)
\nl&& {}\times
\theta\! \left( y_1-\frac{m_k^2}{|2p_jp_k|} x_1 \right) 
\theta\! \left( y_2-\frac{m_l^2}{|2p_jp_l|} x_2 \right).
\end{eqnarray}

Neglecting angular-dependent NNLL of order
$\log^2{(s/M^2)}\log^2{(2p_mp_n/s)}$, we obtain the following results
for the cases \refeq{smcase2} in \NLLad\ approximation:
\beqar
\lefteqn{
\I_{3\rL} (\lambda,\lambda;p_j,p_k,p_l) \LA
\frac{1}{2} \log \frac{m_j^2m_k^2}{(2p_jp_k)^2} \log \frac{m_j^2m_l^2}{(2p_jp_l)^2} \log^2
 \frac{\lambda^2}{|2p_jp_k|} 
}\quad&&
\nl &&{}
- \frac{1}{2} \log \frac{m_j^2m_k^2}{(2p_jp_k)^2} \log
 \frac{\lambda^2}{|2p_jp_k|}
 \left[ \log^2 \frac{m_j^2}{|2p_jp_l|}+\log^2 \frac{m_l^2}{|2p_jp_l|}
\right.\nl&&{}\qquad\left.
+ \log \frac{m_j^2m_l^2}{(2p_jp_l)^2}
 \left( \log \frac{m_k^2}{m_l^2} - 4 \log \frac{|p_jp_k|}{|p_jp_l|}\right) \right]
\nl&&{}
+ \frac{1}{8} \left[ \log^2 \frac{m_j^2}{|2p_jp_l|}+\log^2 \frac{m_l^2}{|2p_jp_l|} \right]
\left[ \log^2 \frac{m_j^2}{|2p_jp_k|}+\log^2 \frac{m_k^2}{|2p_jp_k|} \right]
\phantom{g^4 \frac{M^2}{|2p_jp_k|}}
\nl&&{}
+\log^3 \frac{m_j^2}{|2p_jp_k|}\left[ \frac{2}{3} \log
\frac{m_k^2}{m_l^2}- \frac{19}{12}\log\frac{|p_jp_k|}{|p_jp_l|} 
\right] 
\nl&&{}
+ \log^2 \frac{m_j^2}{|2p_jp_k|}\left[ \frac{3}{8} \log^2
\frac{m_k^2}{|2p_jp_k|}-\frac{3}{8}\log^2\frac{m_l^2}{|2p_jp_k|} 
- \frac{9}{4}\log\frac{|p_jp_k|}{|p_jp_l|} \log\frac{m_l^2}{|2p_jp_k|}
\right] 
\nl&&{}
+\log \frac{m_j^2}{|2p_jp_k|}\biggl[ \frac{1}{6} \log^3\frac{m_k^2}{|2p_jp_k|}
- \frac{1}{6}\log^3\frac{m_l^2}{|2p_jp_k|}
\nl&&{}\qquad
- \log\frac{|p_jp_k|}{|p_jp_l|}
\biggl(\frac{1}{4}\log^2\frac{m_k^2}{|2p_jp_k|}
+\log^2\frac{m_l^2}{|2p_jp_k|}\biggr)
\biggr] 
\nl&&{}
-\frac{1}{24}\log^4 \frac{m_l^2}{|2p_jp_k|}
+\frac{1}{6} \log^3\frac{m_k^2}{|2p_jp_k|}\log\frac{m_l^2}{|2p_jp_k|}
- \frac{1}{8} \log^2\frac{m_k^2}{|2p_jp_k|}\log^2\frac{m_l^2}{|2p_jp_k|}
\nl&&{}
+ \log\frac{|p_jp_k|}{|p_jp_l|}
\left(\frac{1}{3}\log^3\frac{m_k^2}{|2p_jp_k|}
-\frac{5}{4}\log^2\frac{m_k^2}{|2p_jp_k|} \log\frac{m_l^2}{|2p_jp_k|}
-\frac{1}{3}\log^3\frac{m_l^2}{|2p_jp_k|}
\right) 
\nl&&{}
+\theta\left(m_k(p_jp_l)^2-m_l(p_jp_k)^2\right)
\left(\frac{1}{24}\log^4\frac{m_l^2}{m_k^2}
+\frac{1}{3}\log\frac{|p_jp_k|}{|p_jp_l|}\log^3\frac{m_l^2}{m_k^2}\right)
,
\label{eq:s23ll}
\\
\lefteqn{
\I_{3\rL} (M,\lambda;p_j,p_k,p_l) \LA
\left[\frac{2}{3}\log \frac{m_l^2}{|2p_jp_l|}
+\frac{1}{6} \log\frac{M^2}{|2p_jp_l|} \right] \log^3 \frac{M^2}{|2p_jp_l|},
}\quad&&
\label{eq:s23ml}
\\
\lefteqn{
\I_{3\rL} (\lambda,M;p_j,p_k,p_l) \LA \log \frac{\lambda^2}{|2p_jp_k|}
\log \frac{m_j^2m_k^2}{(2p_jp_k)^2} \log^2 \frac{M^2}{|2p_jp_l|} - \frac{1}{6} \log^4 \frac{M^2}{|2p_jp_k|}
}\quad&&
\nl&&{}
-\frac{1}{2} \log^2 \frac{M^2}{|2p_jp_l|} \left( \log^2 \frac{m_j^2}{|2p_jp_k|}+\log^2 \frac{m_k^2}{|2p_jp_k|}
\right)
\nl&&{}
- \frac{2}{3} \log^3 \frac{M^2}{|2p_jp_k|}
\log \frac{m_k^2}{|2p_jp_k|},
\label{eq:s23lm}
\\
\lefteqn{
\I_{3\rL} (M,M;p_j,p_k,p_l) \LA
\frac{1}{2} \log^4
\frac{M^2}{|2p_jp_l|}.
}\quad&&
\label{eq:s23mm}
\eeqar

\subsubsection*{3-leg Yang--Mills diagram $\I_{3\rY}$}
For the 3-leg Yang--Mills diagram 
\beqar\label{3legsym}
\diagthreeY{jkl}{abc}:=
\vcenter{\hbox{
\eikdiagramthreeY
}} 
\quad
&\EIK&
\begin{array}{l}
\\ \displaystyle{
\left(\frac{\alpha}{4\pi}\right)^2
\I_{3\rY}(M_a,M_b,M_c;p_j,p_k,p_l)
}\\ \displaystyle{\hspace{0mm}{}\times
\M_0^{i_1\dots i'_j\dots i'_k\dots i'_l\dots i_n}
I^{\bar a}_{i'_j i_j}
I^b_{i'_k i_k}
I^{\bar c}_{i'_l i_l}
I^a_{bc}
}\end{array}
\eeqar
we have
\beqar
\lefteqn{\hspace{-0.5cm}
\I_{3\rY} (M_a,M_b,M_c;p_j,p_k,p_l)
:=  (4\pi)^4 \int \frac{\rd^4l_1}{(2 \pi)^4} \int \frac{\rd^4l_2}{(2 \pi)^4}
\frac{1}{ [l_1^2-M_a^2][(p_j-l_1)^2-m_j^2]}
}\quad&&
\nl&&{}\times
\frac{8p_kp_l [(l_1 +2 l_2)p_j]
-8p_jp_k[(l_2-l_1)p_l]
-8p_jp_l[(2l_1+l_2)p_k]}{
[l_2^2-M_b^2][(l_1+l_2)^2-M_c^2] [(p_k-l_2)^2-m_k^2] [(p_l+l_1+l_2)^2-m_l^2]}.
\label{eq:s32}
\end{eqnarray}
Neglecting angular-dependent NNLL of order
$\log^2{(s/M^2)}\log^2{(2p_mp_n/s)}$, we obtain the following results
for the cases \refeq{smcase3} in \NLLad\ approximation:
\beqar
\lefteqn{
\I_{3\rY} (\lambda,M,M;p_j,p_k,p_l) 
=\I_{3\rY} (M,\lambda,M;p_l,p_j,p_k) 
=\I_{3\rY} (M,M,\lambda;p_k,p_l,p_j) 
}\quad&&
\nl
&\LA&
-\frac{1}{3} \log \frac{|p_jp_k|}{|p_jp_l|}
\log^2 \frac{M^2}{|2p_kp_l|} \left[ \log \frac{M^2}{|2p_kp_l|}-3 \log
  \frac{m_j^2}{|2p_kp_l|} \right],
\phantom{M,\lambda;p_k,p_l,p_j}
\label{eq:s32lmm} 
\nl
\lefteqn{
\I_{3\rY} (M,M,M;p_j,p_k,p_l) \LA0.
}\quad&&
\end{eqnarray}

\subsubsection*{4-leg ladder diagram $\I_{4\rL}$}
Finally, for the 4-leg ladder diagram 
\beqar\label{4legsladder}
\diagfourL{jklm}{ab}:=
\vcenter{\hbox{
\eikdiagramfourL
}} 
\quad&\EIK&
\begin{array}{l}
\\\displaystyle{
\left(\frac{\alpha}{4\pi}\right)^2
\I_{4\rL}(M_a,M_b;p_j,p_k,p_l,p_m),
}\\\displaystyle{
\hspace{0mm}{}\times
\M_0^{i_1\dots i'_j\dots i'_k\dots i'_l\dots i'_m\dots i_n}
I^a_{i'_j i_j}
I^{\bar{a}}_{i'_k i_k}
I^b_{i'_l i_l}
I^{\bar{b}}_{i'_m i_m}
}\end{array}
\eeqar
we have
\beqar\label{forulegint}
\I_{4\rL}(M_a,M_b;p_j,p_k,p_l,p_m)&=&
\Ione(M_a;p_j,p_k)
\Ione(M_b;p_l,p_m).
\eeqar


\section{Relations between loop integrals in logarithmic approximation}
\label{se:looprelations}
In the following we list the relations between the one- and two-loop
integrals that have been used in \refse{se:twoloop} in order to
simplify the sum over all eikonal contributions.  These relations have
been obtained from the results of \refapp{se:oneloopint} and
\refapp{se:twoloopint}, and are valid in \NLLad\ 
approximation\footnote{The relations are actually valid for the
  $\theta$-function representations given in \refapp{app:integrals} as
  well, which contain also NNLL.} \refeq{LA}.

Combinations of 2-leg ladder integrals can be expressed as products of
one-loop integrals using
\beqar\label{eq:ladqed}
\left[\I_{2\rL} (M_a,M_b;p_j,p_k) + \I_{2\rC} (M_a,M_b;p_j,p_k)\right] +(a\leftrightarrow b)&\LA&
\Ione (M_a;p_j,p_k)
\Ione (M_b;p_j,p_k), \nl
\eeqar
which is valid for all cases \refeq{smcase2}.
The 2-leg Yang--Mills diagram can be related to the 2-leg crossed
ladder diagram using
\beq\label{eq:2Yid} 
\I_{2\rY} (M_a,M_b,M_c;p_j,p_k) \LA - \frac{1}{2} \I_{2\rC} (M_a,M_b;p_j,p_k), 
\eeq
which is valid in all cases \refeq{smcase3}.
Furthermore, we have
\beqar\label{eq:ladid}
\I_{2\rL} (M,\lambda;p_j,p_k)&\LA& \I_{2\rL} (M,M;p_j,p_k),
\nl
\I_{2\rC} (M,\lambda;p_j,p_k)&=& \I_{2\rC} (\lambda,M;p_k,p_j). 
\eeqar
The  relation 
\beq\label{eq:angladid}
\I_{3\rL} (M_a,M_b;p_j,p_k,p_l) + \I_{3\rL} (M_b,M_a;p_j,p_l,p_k) \LA \Ione(M_a;p_j,p_k) \Ione (M_b;p_j,p_l)
\eeq
permits to simplify combinations of 3-leg ladder integrals for all
cases \refeq{smcase2}.  Furthermore
\beq\label{eq:antisymmperm}
\sum_{\pi(j,k,l)}\sgn(\pi(j,k,l)) \I_{3\rL}(M,M;p_{\pi_l},p_{\pi_k},p_{\pi_j}) \LA 0,
\eeq
where the sum runs over all permutations $\pi(j,k,l)$ of $j,k,l$, and
$\sgn(\pi(j,k,l))$ is the sign of the permutation. For the 3-leg
Yang--Mills diagram we have
\begin{equation}\label{eq:antisymmzero}
\I_{3\rY} (M,M,M;p_j,p_k,p_l) \LA 0,
\end{equation}
as is evident from the totally antisymmetric property in the external
momenta $p_j,p_k$ and $p_l$.  In presence of a photon, this feature is
absent owing to the mass gap $\la\ll M$, and we find instead
\beqar\label{eq:ymid}
\I_{3\rY} (M,\lambda,M;p_j,p_k,p_l) 
&\LA& \frac{1}{2} \Bigg[ \Delta \I_{3\rL} (M,\lambda;p_j,p_l,p_k) 
- \Delta \I_{3\rL} (M,\lambda;p_l,p_j,p_k)\Bigg],
\eeqar
where $\Delta \I_{3\rL}$ is the subtracted part of the 3-leg ladder
diagram as defined in \refeq{eq:subtraction}. This is related to the
subtracted part of the 2-leg crossed ladder diagram by
\beq\label{eq:3L2Cid}
 \Delta \I_{3\rL} (M,\lambda;p_j,p_l,p_k) 
\LA \Delta \I_{2\rC} (M,\lambda;p_j,p_k).
\eeq

\end{appendix}

\end{document}